\documentclass[%
 reprint,
 amsmath,amssymb,
 aps,
]{revtex4-2}

\usepackage{graphicx}
\usepackage{dcolumn}
\usepackage{bm}
\usepackage[version=4]{mhchem} 
\usepackage{placeins} 
\usepackage{float}
\usepackage{graphicx}
\usepackage{multirow} 
\usepackage{siunitx} 

\begin{document}

\preprint{APS/123-QED}

\title{{Interplay of network architecture and ionic environment \\in dictating pNIPAM microgel thermoresponsiveness}}

\author{Syamjith KS}
\author{Alan Ranjit Jacob}%
 \email{arjacob@che.iith.ac.in}
\affiliation{Soft Matter Group, Department of Chemical Engineering, \\ Indian Institute of Technology, Hyderabad, India
}
\date{\today}

\begin{abstract}

The utility of non-functionalized poly(N-isopropylacrylamide) (pNIPAM) microgels in physiological and environmental applications is strictly dependent on their reversible thermoresponsiveness and stability in saline media. Despite their importance, a unified understanding of how network topology, specifically crosslinker concentration and distribution regulates ionic sensitivity remains fragmented in the literature. This work systematically investigates the interplay between network topology and ionic strength (0–100 $mM$ $NaCl$) across eight distinct microgel architectures, ranging from ultra-low crosslinked (ULC) to core-corona and homogeneously crosslinked (HC) variants. Utilizing dynamic light scattering across 22 batches, we analyzed critical thermoresponsive properties, including volume phase transition temperature (VPTT) shifts, salt tolerance thresholds, hysteresis indices, and flocculation kinetics (only at extreme salinity, 1000 $mM$ $NaCl$ and at 25$^{\circ}C$). This comprehensive investigation enables a multi-dimensional analysis of how ionic strength, the presence or absence of crosslinkers (MBA), spatial crosslinking distribution, and thermodynamic states dictate microgel behavior across varying temperatures. Finally, we evaluate the applicability of this experimental library to established theoretical frameworks, specifically the Flory-Rehner and Flory-Rehner-Donnan models, addressing ongoing debates regarding their validity in describing complex microgel systems.

 \end{abstract}

\maketitle

\section{Introduction}

Microgels are soft, crosslinked polymer colloidal particles with a size range from 100$nm$ to 10$\mu m$. They are well known for their unique ability to swell up and collapse back in good solvent\cite{scheffold2020pathways,franco2025soft,ks2025revisiting,tavagnacco2025thermoresponsive,liu2016thermo,nothdurft2021microgel}. Poly(N-isopropylacrylamide), or pNIPAM, is a classic example of a thermoresponsive polymer that is characterized by a lower critical solution temperature (LCST)\cite{winning2024thermoresponsiveness,buratti2022role}. While pNIPAM polymer is crosslinked into a microgel particle, it retains the thermoresponsive behavior, showing a characteristic volume phase transition temperature (VPTT)\cite{morris1997adsorption,fernandez2009gels,nishizawa2025interfacial,perez2015anions,buratti2025copolymer,lopez2007macroscopically,zha2007synthesis,winning2024thermoresponsiveness, bocanegra2022crystallization,sennato2021double,nishizawa2022clarification}, where its properties change significantly. Since its first synthesis by Pelton and Chibante in 1986 \cite{pelton1986preparation}, these systems have been widely studied due to their reversible thermoresponsiveness. In this study, we focus specifically on non-functionalized pNIPAM microgels—defined here as architectures synthesized without the incorporation of ionic comonomers—to isolate the role of network topology  on their fundamental behavior.

\subsection{\label{sec:level8}pNIPAM microgel network architecture and composition effects on properties}
Crosslinking density is a crucial parameter that influences the microstructure and thermoresponsive properties of pNIPAM microgels. There is a plethora of data available in the literature regarding the relationship between crosslinking density and the size, charge, and mechanical properties of microgels with respect to temperature\cite{kratz2001structural, parasuraman2012poly, rumyantsev2014theory,kawaguchi2020going}. In our recent work \cite{ks2025revisiting}, building on this literature, we synthesized microgels by systematically varying the concentration of crosslinker (N, N'-methylenebisacrylamide, MBA, or BIS) during synthesis to create microgels with different crosslinking densities. The results show that increasing the MBA concentration decreases the hydrodynamic diameter in the swollen state and elevates both the VPTT and electrokinetic transition temperature (ETT). These microgels exhibited reversible thermoresponsive behavior with minimal hysteresis irrespective of the crosslinking density.

 Microgels can be synthesized in the absence of crosslinkers, and these are referred to as the ultra-low crosslinked (ULC) microgels\cite{scotti2019exploring,bachman2015ultrasoft}. Earlier studies by Jun et al.\cite{gao2003cross,gao2005influence} established that ULC microgels formed through self-crosslinking chain transfer reactions \cite{hu2011control} during free radical polymerization, exhibits volume phase transition (VPT) near 32–33$^{\circ}C$.  Otto et al.\cite{virtanen2016next} reported an inverted polymer density profile in ULC microgels, where crosslinking dominates near the surface of the particle rather than the core. However, Dimitri et al.\cite{wilms2021elastic} utilized the AFM nanoindentation mapping method to show that self-crosslinked microgels display a more homogeneous elastic profile compared to microgels synthesized with the crosslinker (which exhibit a core-corona structure). 

In addition to the crosslinking density, the distribution of crosslinking throughout the microgel also impacts its properties. These conventional microgels have a dense, highly crosslinked core and loose corona due to faster reaction kinetics for crosslinking (MBA) than polymerization (NIPAM)\cite{ks2025revisiting}.  In contrast, Homogeneous Crosslinked (HC) microgels have a uniform crosslinking distribution throughout the particle. The semi-batch surfactant-free emulsion polymerization method\cite{still2013synthesis,kwok2013controlling},in which the crosslinker is fed at regular intervals during the reaction, is expected to reduce heterogeneity in the crosslinking distribution within a microgel. Saisavadas et al. \cite{saisavadas2023large} with the help of  large-amplitude oscillatory shear rheology, showed that dense HC microgels yield like glasses in a single step, unlike core-corona microgels, which yield in two steps due to entangled chains in the loose corona. Together, these studies underscore that the crosslinking density and its distribution influence microgel's properties.

\subsection{\label{sec:level3}pNIPAM microgel behavior in salt environments}
Beyond temperature, the chemical environment—specifically the ionic strength—plays an important role in the microgel thermodynamics.  It can be analyzed through two distinct aspects, the Debye-Hückel electrostatic screening\cite{denton2016counterion,quesada2014temperature,zha2002effect,lee2015effect,yang2020temperature,kosovan2015modeling,khan2013preparation,shao2013role,sean2017computer} and the salting-out effect\cite{feher2019effect,gan2009situ,lopez2006cationic,karg2008temperature,kokufuta1998volume,ohmura2025osmosis,bischofberger2015new,tan2010microstructure,fanaian2012effects,lee2015effect,kobayashi2017polymer,burba2008salt,moghaddam2019hofmeister,sean2017computer}. These two mechanisms lead to a decrease in the stability of the microgel suspension, which initiates flocculation \cite{ji2024sustainable,daly2000study,ma2006flocculation,daly2000study,rasmusson2004flocculation,rasmusson2004flocculation1,liao2012fractal,hu2011synthesis,adroher2016effect,liu2008cationic,perez2018effect,routh2002salt}. Daly and Saunders\cite{daly2000study} reported that anions with high charge density (citrate$^{3-}$) lead to dehydration and then flocculation. They reported the critical flocculation temperature (CFT), the temperature above which irreversible particle aggregation begins, often coincides with the LCST and shifts systematically with salt concentration and type of salts. The critical flocculation concentration (CFC), defined as the minimum salt concentration required to trigger flocculation, it follows the Schulze-Hardy rule, with higher-valence cations exhibiting greater screening efficiency\cite{zhu2024cation} thereby inducing flocculation at lower concentrations. As reported in the literature\cite{lopez2007macroscopically, lopez2007hofmeister,perez2018effect}, kosmotropic anions like Cl$^{-}$ and F$^{-}$ in the Hofmeister series trigger more pronounced deswelling as compared to chaotropic anions like SCN$^{-}$\cite{pastoor2012anion,lopez2010salt}, and reducing the transition temperature of pNIPAM microgels by disrupting its hydration layer. With the help of classical molecular dynamics simulations\cite{du2010effects} and using high-resolution nuclear magnetic resonance (NMR) spectroscopy\cite{pastoor2015cation}, it has been reported that cations (Na$^{+}$, K$^{+}$, Ca$^{+}$, Mg$^{2+}$) interact directly with the neutral pNIPAM polymer through ion-dipole interactions with carbonyl oxygen atoms, and depressing the LCST. The depression in the transition temperature is by altering both enthalpic and entropic contributions to the phase transition, with the magnitude of depression reflecting the ion's position in the Hofmeister series\cite{guan2011pnipam,fussell2023controlling,quesada2014thermo,perez2015anions,lopez2014thermally}. It is important to note that, while NaCl is positioned in the middle of the Hofmeister series, even NaCl and other intermediate salts can significantly affect thermoresponsive polymer behavior\cite{zhu2024ionic}.

\subsection{\label{sec:level4}Thermoreversibility and hysteresis}
Thermoreversibility and hysteresis are critical that define the functional stability of pNIPAM microgels, reflecting their ability to undergo repeated volume phase transitions during heating–cooling cycles without significant loss of function\cite{pan2018superfast,hoare2007calorimetric,brijitta2008phase,vdovchenko2021effect,hu2004hydrogel}. Hai et al. (2025) \cite{hai2025poly} reported that pNIPAM microgel-based hydrogels maintained a reversible size reduction with shape recovery ratios exceeding 90\% in the presence of high NaCl concentrations. However, Li-Li Zhao et al. (2025)\cite{zhao2025temperature} noted that excessive ionic content is beneficial for thermochromic properties, but it can affect mechanical properties. Yuichiro et al. (2024) \cite{nishizawa2025determination} reported that deswelling occurs by a stepwise core-to-shell sequence, where shifts in softness and charge density generate history-dependent hysteresis during thermal cycling. Ionic strength significantly impacts volume phase transitions, as seen in poly(acrylamide-co-sodium acrylate) hydrogels\cite{okay2000swelling}, where increasing NaCl concentrations reduce swelling due to Donnan osmotic pressure\cite{nishizawa2022clarification,hirotsu1987volume,lopez2007macroscopically,fernandez2000charge}. 

\subsection{\label{sec:level5}The microgel swelling model}

The degree of salt-induced deswelling depends on solvent quality, which can be explained through the Flory-Huggins interaction parameter ($\chi$)\cite{bordi2003electrical,hoare2007functionalized}. Researchers have been attempting to employ the classical Flory-Rehner model to describe the equilibrium swelling behavior of pNIPAM microgels. The equilibrium state of a microgel is defined by the point where the total osmotic pressure is equal to zero (Eq. \eqref{eq:flory_rehner}).  The two competing forces in equilibrium swelling are first, the mixing term, which indicates swelling of the polymer chains in the solvent; this is governed by the Flory-Huggins interaction parameter ($\chi$).  The second term is the elastic term that acts as a restoring force, pulling the network back, meaning that in a more densely crosslinked microgel the polymer chains collapse more drastically.

\begin{figure*}[t]
\centering
\begin{equation}
\label{eq:flory_rehner}
\underbrace{
    \ln\left(1-\left(\frac{R_{\text{H},0}}{R_{\text{H}}(T)}\right)^3\phi_0\right) + \left(\frac{R_{\text{H},0}}{R_{\text{H}}(T)}\right)^3\phi_0 + \left(\frac{R_{\text{H},0}}{R_{\text{H}}(T)}\right)^6\phi_0^2\chi
}_{\text{Mixing term}} 
+ \underbrace{
    \frac{\phi_0}{N_{\text{Seg}}}\left[\frac{R_{\text{H},0}}{R_{\text{H}}(T)}-\frac{1}{2}\left(\frac{R_{\text{H},0}}{R_{\text{H}}(T)}\right)^3\right]
}_{\text{Elastic term}} = 0
\end{equation}

\vspace{5pt}

\begin{equation}
\label{eq:phi_relation}
    \phi(T) = \phi_0 \left( \frac{R_{\text{H},0}}{R_{\text{H}}(T)} \right)^3
\end{equation}

\vspace{5pt}

\begin{equation}
\label{eq:hill_model}
\chi(T) = \chi_0 + a(T - T_a) + b \frac{T_{\text{rel}}^v}{T_{\text{rel}}^v + K} \quad \text{where} \quad K = \left( \frac{VPTT - T_a}{T_e - T_a} \right)^v
\end{equation}

\vspace{5pt}

\begin{equation}
\label{eq:frd_model}
\underbrace{
    \ln\left(1-\left(\frac{R_{\text{H},0}}{R_{\text{H}}(T)}\right)^3\phi_0\right) + \left(\frac{R_{\text{H},0}}{R_{\text{H}}(T)}\right)^3\phi_0 + \left(\frac{R_{\text{H},0}}{R_{\text{H}}(T)}\right)^6\phi_0^2\chi
}_{\text{Mixing term}} 
+ \underbrace{
    \frac{\phi_0}{N_{\text{Seg}}}\left[\frac{R_{\text{H},0}}{R_{\text{H}}(T)}-\frac{1}{2}\left(\frac{R_{\text{H},0}}{R_{\text{H}}(T)}\right)^3\right]
}_{\text{Elastic term}} + \underbrace{v_s(c_s^{in} - c_s^{out})}_{\text{Donnan term}} = 0
\end{equation}

\vspace{10pt}
\small
\begin{tabular}{llll}
   $R_{\text{H}}(T)$,  $R_{\text{H},0}$ & Hydrodynamic radius at temp $T$ and collapsed state & $T_a, T_e$ & First and last temperature data points  \\
    $a$ & Slope of the baseline & $T_{\text{rel}}$ & Relative temperature $(T-T_a)/(T_e-T_a)$  \\
    $\phi(T)$ & Polymer volume fraction of a microgel at temp $T$ & $b$ & Amplitude parameter of Hill transition  \\
    $\phi_0$ & Polymer volume fraction of a microgel at collapsed state & $v$ & Hill parameter (cooperativity) \\
    $\chi_0$ & Interaction parameter at $T_a$ & $VPTT$ & Volume phase transition temperature  \\
    $N_{\text{Seg}}$ & Chain length between crosslinks & $v_s$ & Molar volume of solvent \\
    $c_s^{in}$ & Internal ion concentration & $c_s^{out}$ & External ion concentration \\
\end{tabular}

\hrulefill 
\end{figure*}

While the Flory-Rehner framework is the cornerstone of polymer gel swelling physics, its application to microgel is currently a subject of debate in the literature. The debate centers primarily around the physical validity of the parameters in the model. In 2017, Lopez and Richtering \cite{lopez2017does}, highlighted shortcomings of the classical approach by applying the Flory-Rehner framework to the data set of the pNIPAM microgel systems. They argued that assuming an affine network (the network deforms exactly like the macroscopic gel) overestimates the elastic forces, and they evaluated that the $\phi_0$ (polymer volume fraction in the collapsed state) of the microgel should be physically limited to around 0.44, which differs from the reported best-fit value of 0.8. $\phi_0$ is the polymer volume fraction in the collapsed state of the microgel, thereby providing an estimate of the amount of polymer that remains in the collapsed microgel after the water is expelled during the VPT. Lopez and Richtering reported that the classical value of $0.8$ for $\phi_0$ is not valid in the case of pNIPAM microgel. However, in 2018, DC Leite et al.\cite{leite2018smart}, defended the classical framework by employing it for starch-pNIPAM hybrid microgels. They reported $\phi_0$ as 0.8 for successful fits, while force fitting $\phi_0$ as 0.44 was found to be unsuccessful. The discrepancy in $\phi_0$ values arises from the estimation of the microgel's molar mass, which is determined by Lopez and Richtering \cite{lopez2017does}. In addition to that, DC Leite et al.\cite{leite2018smart}, used two equations for $\chi$, (i) a series expansion up to $\phi^3$ and (ii) a Hill-like model(Eq. \eqref{eq:hill_model}). The Hill-like approach was found to better represent the cooperative thermotropic phase transition. Later in 2020, Lopez et al.\cite{lopez2020electrostatic} fit the Flory-Rehner-Donnan model (Eq. \eqref{eq:frd_model}) for the ionic microgel in different ionic strengths of the  medium. Thus, introducing  the Donnan term in the framework\cite{katchalsky1955polyelectrolyte}, which accounts for the osmotic pressure generated by the concentration gradients of the mobile counterions within the microgel and bulk medium. Using $\phi_0$ as 0.44 the model fails unless $N_{\text{Seg}}$ is artificially changed to unphysical values at low ionic strengths. In 2022, Friesen et al\cite{friesen2022modified}, employed a modified version of the Flory-Rehner theory for copolymer microgels. Their experimental fits yielded $\phi_0$ as 0.7-0.89. More importantly, they restricted the model to just 3 adjustable parameters and argued that the model can still capture essential physics.

Although the fundamental thermoresponsive properties of pNIPAM microgel are well-documented, the complex interplay between specific microgel architectures and their stability in varying ionic environments remains fragmented. To address the gaps in the current understanding of microgel physics, this work is structured around two core objectives. Initially, the evaluation of the interplay of network architecture, initiator concentration, and ionic strength that dictates the fundamental thermoresponsive behavior, specifically the changes in the VPTT, swelling capacity, salt tolerance, reversibility of the transition, and flocculation kinetics analysis. Secondly, this manuscript assesses the applicability and physical consistency of Flory-Rehner and Flory-Rehner-Donnan models to determine whether classical parameters capture the underlying physics across diverse networks of architectures and experimental conditions. 

The rest of this manuscript is structured to address these objectives. Section 2 details the protocols used to synthesize the architecturally distinct pNIPAM microgels, then describes the advanced characterization techniques employed to study their thermoresponsive and colloidal behavior, and finally applies the data sets to the classical models. In section 3, the results and analysis, covering the effects of crosslinking density and distribution and initiator concentration on VPTT shifts, salt tolerance, flocculation kinetics, thermal hysteresis, and model fits to the experimental data, are presented. Finally, section 4 provides a summary of key findings and discusses their research implications.

\section{Materials and Methods}
\subsection{\label{sec:level6}Chemicals}
All reagents were purchased from Sigma Aldrich and used without further treatment. NIPAM (\ce{C6H11NO} $>$98\%), KPS (\ce{K2S2O8}, $>$97\%), MBA (\ce{C7H10N2O2}, $>$99\%), SDS (\ce{NaC12H25SO4}, $>$98\%), and \ce{NaCl} ($>$99\%) were used in the synthesis and characterizations. The chemical structures of NIPAM and MBA are provided in the  supplemental material (Fig. S1) to illustrate the free radical polymerization and crosslinking mechanisms during microgel synthesis. Ultrapure deionized water (type 1/milli-Q, resistivity 18 $\text{M}\Omega\cdot \text{cm}^{-1}$) used as the solvent. Prior to use, type 1 water was filtered through a \SI{0.4}{\micro\meter} syringe filter to remove impurities.
\subsection{\label{sec:level7}Microgel synthesis}
Microgels with different network architectures and compositions were synthesized using two synthesis methods (conventional one-pot and semi-batch synthesis), as discussed in our earlier work \cite{ks2025revisiting}. A detailed overview of the composition and synthesis setup of two types of synthesis is provided in the supplementary information\cite{}. In total, 8 different microgel formulations were synthesized, of which 7 with different crosslinking densities (including the ULC), and 1 with homogeneous crosslinking. Multiple replicates of each variant were prepared, resulting in a total of 22 batches. This set of batches enabled rigorous analysis and validation of microgel characterization.

\subsubsection{Conventional one-pot synthesis}
pNIPAM microgel with a core-corona microstructure was prepared using conventional one-pot synthesis. This approach allowed systematically changing the composition during syntheses, including the concentration of crosslinker and initiator, with all other reaction parameters held constant. The syntheses yielded 7 types of microgels with different MBA/NIPAM mole ratios (0.007, 0.031, 0.038, 0.058, 0.077, and 0.097), along with one ULC microgel synthesized without MBA\cite{gao2003cross,gao2005influence,brugnoni2019synthesis,babenyshev2025size}. 

\subsubsection{Semi-batch synthesis}
Semi-batch synthesis was utilized to prepare a pNIPAM microgel with homogeneous crosslinking density (HC) with an MBA/NIPAM mole ratio of 0.020. In semi-batch synthesis, a feed solution of monomer and crosslinker is added gradually during the reaction, allowing more controlled network formation and reduced heterogeneity in crosslinking distribution\cite{saisavadas2023large,still2013synthesis}.

Detailed synthesis protocols, including reagent composition and reaction conditions, are explained in the supplementary material.

\subsubsection{Reaction Mechanism}
The reaction mechanism during the synthesis of pNIPAM microgels involves free radical polymerization initiated by KPS. The NIPAM monomer has an isopropyl group, an amide group, and a reactive vinyl group. The vinyl group is the primary site for chain propagation. The initiator, KPS, undergoes thermal decomposition upon addition to the reaction mixture at 70$^{\circ}C$, generating active sulfate radical ($SO_4^{\bullet-}$) species. These free radicals initiate polymerization by attacking the vinyl groups of the NIPAM monomers, leading to the growth of linear pNIPAM chains. MBA, the crosslinker, acts as the bifunctional comonomer by crosslinking these pNIPAM chains to facilitate the formation of a stable three-dimensional network. Because MBA contains two reactive vinyl groups, it acts as a covalent bridge between two growing polymer chains\cite{wilms2021elastic}. As previously noted, simultaneous polymerization and crosslinking dictate the core-corona architecture of the microgels. This structure consists of a highly crosslinked core and a loosely crosslinked corona, as a result of the faster reaction kinetics of the crosslinker (MBA) relative to the monomer (NIPAM)\cite{ks2025revisiting}. The reaction mechanism and structural schematic are illustrated in the supplementary material (Fig. \ref{SI2}).

The ULC microgels are formed in the absence of MBA, where microgel network formation occurs primarily through self-crosslinking and chain transfer reaction mechanisms. The self-crosslinking happens by abstracting hydrogen from the isopropyl group's tertiary carbon\cite{brugnoni2019synthesis,babenyshev2025size}. A detailed schematic of the chemical structure and formation mechanism of ULC microgels is provided in the supplementary material (Fig. \ref{SI10}).

\subsection{\label{sec:level9}Dynamic light scattering (DLS) for size analysis in varying ionic strength medium}
DLS was employed to investigate the hydrodynamic diameter of synthesized pNIPAM microgels across varying ionic strength environments and temperatures. All pNIPAM microgel samples mentioned in this manuscript were prepared at a concentration of 0,33 $mg/mL$. The ionic strength of the medium was varied by adjusting $NaCl$ concentrations (0 (pure water), 0.1, 1, 10, and 100 ($mM$). After the addition of $NaCl$ to the microgel suspensions, samples were mixed for a minimum of 4 hours using a roller mixer to ensure homogeneous mixing.

The spherical morphology of the synthesized microgels was confirmed by atomic force microscopy (AFM), as described in our previous work\cite{ks2025revisiting}. DLS measurements are particularly well-suited for characterizing the size of spherical particles in dispersion. Anton Paar particle analyzer (LiteSizer 500) equipped with a 658 $nm$ laser light source was used for all measurements. The instrument was operated at a 90\textsuperscript{o} scattering angle (side scattering) to optimize the signal-to-noise ratio.
To assess the reversible thermoresponsive behavior of pNIPAM microgels, size measurements were performed during both heating and cooling cycles. During the heating cycle, the temperature was systematically varied from 20$^{\circ}C$ to 50$^{\circ}C$ with step increments of 0.5$^{\circ}C$. We collected 61 points per sample per heating cycle. The temperature was subsequently decreased from 50$^{\circ}C$ back to 20$^{\circ}C$ using step increments of 2.5$^{\circ}C$ during the cooling cycle. At each temperature increment, the microgel sample was allowed to equilibrate for 10 min to ensure that the system reached thermal equilibrium\cite{ks2025revisiting}.

DLS data were analyzed using the cumulant approach. At high $NaCl$ concentrations and elevated temperatures, large particle sizes indicated flocculation. Therefore, data points where the microgel sizes at elevated temperatures were greater than the size of microgel in pure water were excluded.

Estimation of the characteristic VPTT is crucial for pNIPAM microgels. The VPTT is the temperature at which the microgel hydrodynamic diameter exhibits a significant reduction\cite{ks2025revisiting}. Following DLS temperature sweep experiments, size data were obtained across a range of temperatures and ionic strength of the medium. As mentioned above, at high $NaCl$ concentrations and elevated temperatures, the data points of flocculation were cropped even during the fitting. The VPTT extracted from the size-temperature data; a phenomenological function analysis was employed\cite{saunders2004structure, hertle2013thermoresponsive} a mathematical function to fit a sigmoid curve to experimental data as detailed in our previous work\cite{ks2025revisiting}.

{\subsection{\label{sec:level10}Introduction of indexes: Normalized size and reversibility index}

\subsubsection{Normalized size ($\lambda_D$) for salt tolerance analysis}}
To quantify the salt tolerance of pNIPAM microgels across various $NaCl$ concentrations, a scaled metric, normalised size ($\lambda_D$), was introduced. $\lambda_D$ is calculated as the ratio of the size of microgel in a salt solution to the size of microgel in pure water:

\begin{equation}
\text{$\lambda_D$} = \frac{D_{\text{h, salt}}}{D_{\text{h, water}}}
\end{equation}

Where,

$D_{\text{h, salt}}$: Size of microgel in the salt solution

$D_{\text{h, water}}$: Size of the microgel in pure water. 

$\lambda_D$ enables quantitative assessment of microgel size changes in response to $NaCl$ concentration in the medium. $\lambda_D$ values are analyzed across three different thermodynamic states: swollen state (below VPTT, averaged over 20-25$^{\circ}C$), transition (at VPTT of the microgel in pure water, 0 $mM$ $NaCl$), and collapsed state (above VPTT, averaged over 45-50$^{\circ}C$). $\lambda_D$ $\approx 1$ indicates the ideal case, high salt tolerance with minimal size change with salt addition. Deviation from 1 reflects size changes due to the presence of salt. $\lambda_D < 1$ indicates salt-induced deswelling (salting-out), whereas $\lambda_D > 1$ suggests salt-induced swelling (salting-in) and $\lambda_D > 1.5$ implies flocculation.

\subsubsection{Hysteresis index (HI) for thermoreversibility quantification}
To quantify the thermoreversibility and thermal hysteresis of microgels at different $NaCl$ concentrations, the hysteresis index (HI) was defined. The HI is calculated from the area enclosed between the heating and cooling size curves during temperature sweep experiments:

\begin{equation}
\text{HI} = \frac{\left|A_{Heating} - A_{Cooling}\right|}{A_{Heating}} \times 100
\end{equation}

Where, 

$A_{Heating}$: Area under the heating curve

$A_{Cooling}$: Area under the cooling curve

The areas under the curves are determined by numerically integrating the hydrodynamic diameter ($nm$) with respect to temperature ($^{\circ}C$) using the trapezoidal rule in Python. This approach captures the change in size with temperature during heating and cooling cycles in a particular environment for a specific microgel, providing a complete picture of the hysteresis. A higher $HI$ value indicates pronounced hysteresis and reduced reversibility, while a lower $HI$ value indicates good thermoreversibility with minimal difference between heating and cooling cycles. Since, $HI$ is a dimensionless quantity, comparisons can be made across different microgel formulations and ionic strength conditions of the medium.

\subsection{\label{sec:level11}DLS for flocculation kinetics}

Flocculation kinetics, the rate at which pNIPAM microgels flocculate in an extreme saline environment, is a crucial aspect to analyze from an application standpoint\cite{sato2025structural,bocanegra2022crystallization,liao2011gelation,mailer2015particle,frisken2001revisiting}. To understand how to conduct a flocculation kinetics study, various studies on different systems have been reviewed.  Early kinetic studies using DLS \cite{tanaka1979kinetics} revealed that once repulsion is suppressed, aggregation proceeds through diffusion-limited mechanisms\cite{routh2002salt}. Mohraz et al.\cite{mohraz2006gelation} established a DLS-based framework that monitors the first cumulant (representing the z-average diffusion coefficient and thus the size) alongside the nonergodicity parameter ($g_2$(0)-1) during electrolyte-induced aggregation. Cipelletti et al.(2000)\cite{cipelletti2000universal} used multispeckle DLS to demonstrate that particle aggregation kinetics exhibit non-exponential behavior, with relaxation dynamics becoming progressively slower as the gel network ages through internal restructuring mechanisms. With the help of differential dynamic microscopy, Cho et al. (2020)\cite{cho2020emergence}, revealed  distinct temporal stages in colloidal gel formation. Wu et al. (2013)\cite{wu2013kinetics} investigated colloidal gelation in the reaction-limited cluster aggregation regime using small-angle light scattering and established that, when a dimensionless time accounting for particle concentration and stability is employed, aggregation kinetics from systems with different salt concentrations and particle sizes collapse onto a single master curve.  

pNIPAM microgels undergo flocculation in high $NaCl$ concentrations even at room temperature. To analyze how fast flocculation occurs in distinct microgel formulations, DLS measurements were carried out at room temperature under high ionic strength of the medium. The pNIPAM microgel solution was mixed with $NaCl$ to reach final concentrations of 1000 $mM$ $NaCl$ and 0.33 $mg/mL$ of microgel. The mixture was immediately transferred to the DLS instrument, and a series of size measurements were made continuously over 1000 seconds. To minimize artifacts, the instrument temperature (25$^{\circ}C$) was pre-maintained, and measurements started within 3 seconds of mixing to account for experimental errors.
DLS measurements were done using Anton Paar particle analyzer (LiteSizer 500) operated at a scattering angle of 15\textsuperscript{o} and at 25$^{\circ}C$. The flocculation kinetics study was conducted at room temperature, as the microgel form flocculates under extreme saline conditions at this temperature. The choice of forward scattering angle (15\textsuperscript{o}) was critical for probing the large length scales of microgel flocculates. 

The scattering vector, $q$:
\begin{equation}
q = \frac{4\pi n}{\lambda} \sin\left(\frac{\theta}{2}\right)
\end{equation}
Where,

$n$: Refractive index of the solvent

$\lambda$: Wavelength of the incident light

$\theta$: Scattering angle

 The relaxation of density fluctuations in such large flocculates is most observable at small wave vectors, making 15\textsuperscript{o} optimal for observing flocculation kinetics\cite{mohraz2006gelation,cho2020emergence}.

\subsubsection{Autocorrelation function and decay time analysis}
As flocculates are formed over time, this is reflected in the normalized intensity autocorrelation function ($g^{(2)}(t)$)\cite{cipelletti2000universal,mailer2015particle}. A delay in the decay rate of the autocorrelation function means a slowdown of diffusion dynamics due to large flocculates. The normalized intensity autocorrelation function  ($g_{2}/g_{0}$) can be fitted to an exponential decay model to extract decay time. 

\begin{equation}
g_{2}/g_{0} = Ae^{-t/\tau} + C
\label{autocordeacy}
\end{equation}


Where,

$g_{2}/g_{0}$: Normalized intensity autocorrelation function

$A$: Amplitude

$\tau$: Decay time (or relaxation time) 

$t$: Lag time 

$C$: Baseline offset

$\tau$ in Eq. \ref{autocordeacy} directly characterizes the diffusion dynamics of the flocculates.

To ensure the fitting procedure was robust and minimize the impact of experimental noise at long lag times, the data were truncated. Specifically, only data points satisfying the condition, ($g_{2} - 1 > 0.1$) were included in the analysis, ensuring that the fit is calculated toward the key decay signals.

\subsubsection{Normalization by using Brownian time ($\tau_B$) }

In order to remove dependence on the particle size, the $\tau$ was normalized with Brownian time ($\tau_B$). $\tau_B$ represents the characteristic time for a particle to diffuse its own radius. This time scale is given by:

\begin{equation}
\tau_B = \frac{6\pi\eta R^3}{k_B T}
\end{equation}

Where,

$\eta$: Viscosity of the solvent

$R$: Hydrodynamic radius of microgel in pure water

$k_B$: Boltzmann constant $\left(1.381 \times 10^{-23}J/K\right)$

$T$: Temperature

The normalized decay time, defined as $\tau / \tau_B$, is dimensionless and enables comparison of flocculation dynamics across different microgel formulations and ionic strength conditions of the medium, independent of particle size. To characterize flocculation kinetics across all microgel formulations, $\tau / \tau_B$ was plotted as a function of experimental time.

\subsection{\label{sec:level12}Theoretical analysis: Flory-Rehner and Flory-Rehner-Donnan models}
 For the theoretical analysis, we employed a rigorous fitting strategy in Python. The size-temperature data sets for different microgels in different $NaCl$ concentrations are fitted with Flory-Rehner (Eq. \eqref{eq:flory_rehner}) and Flory-Rehner-Donnan (Eq. \eqref{eq:frd_model}) models, using the Hill-like model for $\chi$(Eq. \eqref{eq:hill_model}). To prevent overfitting, a few assumptions and strategies are employed to reduce the number of free parameters.

 \begin{itemize}
    \item $N_{\text{Seg}}$, representing the chain length between crosslinks, is considered a constant characteristic of a particular microgel, assumed that $N_{\text{Seg}}$ is unaffected by ionic strength or temperature. Particularly, it $N_{\text{Seg}}$ is treated as a fitting parameter for the 0 $mM$ $NaCl$ case, and the obtained value is then used as a constant for other $NaCl$ concentrations. 
    \item VPTT estimated from experiments using a phenomenological function. It was estimated for each microgel and each $NaCl$ concentration and treated as a constant value in the models. 
\end{itemize} 
 
The fixed-parameter approach is done to ensure that the models capture real physics rather than mathematically mimicking the data. Other parameters, $\chi_0$, $a$, and $b$ change with $NaCl$ concentrations, so those cannot be fixed for pNIPAM as done by  Friesen et al\cite{friesen2022modified}. 

 It is necessary to check the charge effect due to the residue of KPS on the microgel. That's the reason for including the Donnan term in the model, but the concentration of ions inside the microgel ($c_s^{in}$) can not be estimated as done by Lopez et al.\cite{lopez2020electrostatic}, because the charge arises from the residue of the initiator but not from the charged monomers. Lopez et al.\cite{lopez2020electrostatic} suggest that $c_s^{in}$ is a function of external salt concentration ($c_s^{out}$), and the fraction of charged monomers ($f_c$). Since $c_s^{in}$ is not constant across different NaCl concentrations, it is treated as a fitting parameter. The external ion concentration ($c_s^{out}$) is ranges from 0 to 100 $mM$. For the 0 $mM$ $NaCl$ case, $c_s^{out} = 0$, and $c_s^{in}$ is constrained within $10^{-3}$ to $10^{-1}$ mM (representing the concentration of counterions in water). In the other $NaCl$ concentrations cases, the limits of the $c_s^{in}$ is given as, $10^{-3} mM < c_s^{in} < c_s^{out}$. The fitted $c_s^{in}$ value provides insight into the microgel's charge density.  
\begin{itemize}
    \item If $c_s^{in}$ = $c_s^{out}$, the microgel exhibits very low charge density, behaving like a non-ionic microgel.
    \item If $c_s^{in} < c_s^{out}$, the system possesses fixed charges.
\end{itemize}    
  
 The quality of the fits is evaluated by estimating $(chi)^{2}$ values. The $(chi)^{2}$ is defined as,

 \begin{equation}
(chi)^{2} = \sum_{T_{a}}^{T_{e}} \frac{(D_{H}(T) - D_{H,fit}(T))^{2}}{D_{H,fit}(T)}
\end{equation}

Where,

$D_{H}(T)$: Experimental value for hydrodynamic diameter at temperature T

$D_{H,fit}(T)$: Model predicted value for hydrodynamic diameter at temperature T

The lower the values of $(chi)^{2}$ the better the fits.

\section{Results and discussion}
{\subsection{\label{sec:level1}Combined influence of network architecture, and ionic strength on VPTT}
\begin{figure}[htpb]
\centering
  \includegraphics[height=6cm]{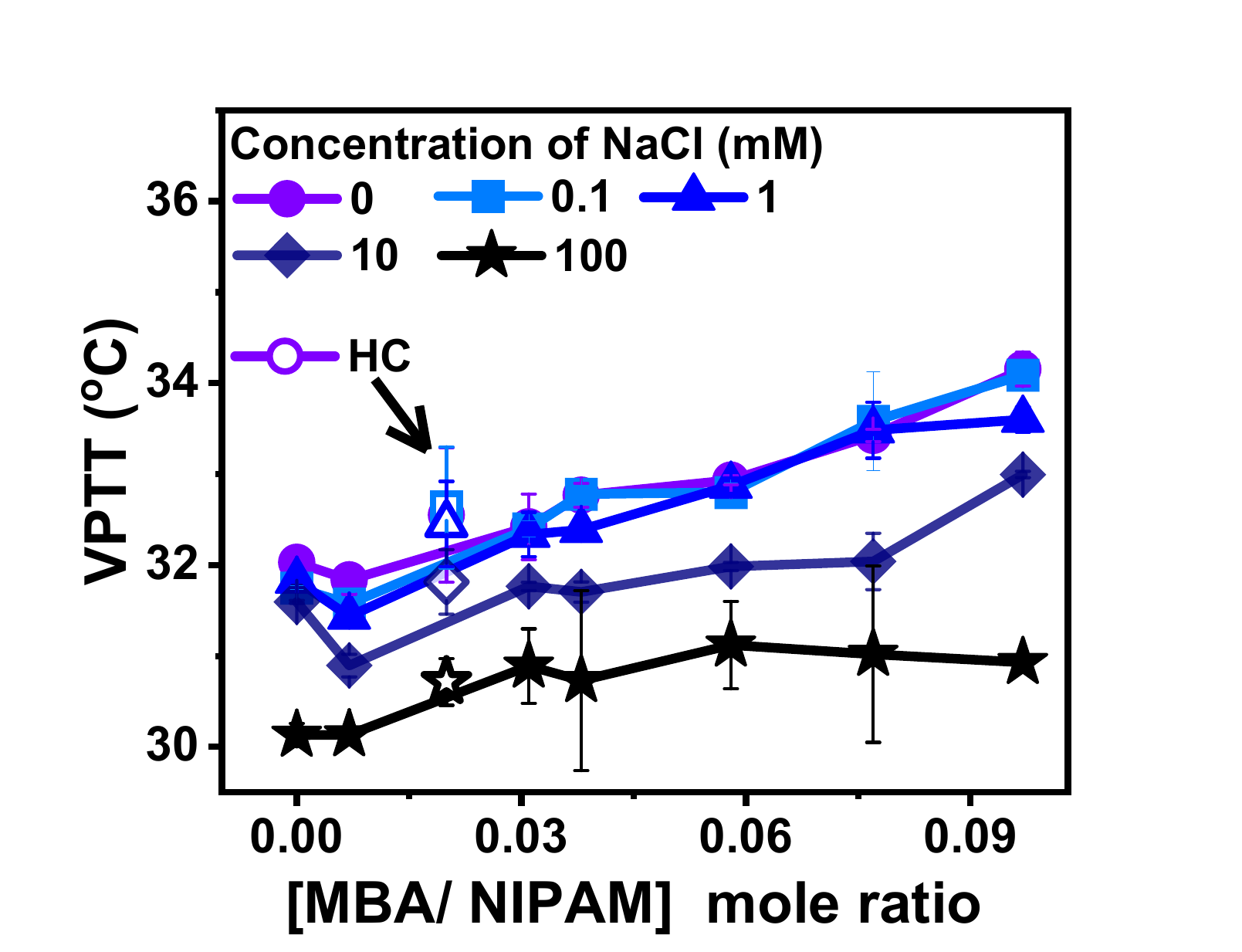}
  \caption{Influence of crosslinker (MBA) concentration and crosslinking distribution (ULC, homogeneous, core-corona). Symbols represent different $NaCl$ concentrations: Circle (0 $mM$, pure water, violet), square (0.1 $mM$, light blue), triangle (1 $mM$,blue), diamond (10 $mM$, navy blue), and star (100 $mM$,black). Closed symbols represent core-corona microgels ([SDS/NIPAM] mole ratio = 0.004, [KPS/NIPAM] mole ratio = 0.016), and open symbols represent homogeneous crosslinking density microgels ([SDS/NIPAM] mole ratio = 0, [KPS/NIPAM] mole ratio = 0.019).}
  \label{fig1}
\end{figure}

VPTT is not determined by a single factor but emerges from a complex interplay of elastic, entropic, and chemical contributions in the microgel. It is the temperature at which the size of the microgel reduces significantly due to the expulsion of water from the microgel network. At room temperature, the water molecules are linked to the network by means of hydrogen bonds with the the amide groups of the monomer and crosslinker of the microgel. The crosslinker, MBA, is a bifunctional co-monomer with two amide groups. Therefore, as crosslinker concentration increases, the network becomes more hydrophilic and exhibits greater elasticity due to a shorter chain length between crosslinks ($N_{\text{Seg}}$). For pNIPAM, as the temperature increases, the solvent-polymer interaction parameter ($\chi$) increases making the polymer-polymer interactions energetically more favorable, leading to the collapse of the microgel. At a molecular level, as the thermal energy of water molecules increases, the hydrogen-bonded hydration shell is disrupted \cite{ks2025revisiting}. To minimize these unfavorable interactions with the water, the pNIPAM chains undergo a coil-to-globule transition, expelling water and leading to the observed shrinkage of the microgel. By taking all of these factors into account, microgels with high MBA concentrations are expected to have higher VPTT, i.e., a high temperature is needed to expel water from the more hydrophilic network.

The salt ions ( $Na^+$ and $Cl^-$) result in the expected salting-out effect. From a molecular perspective, as discussed in the literature\cite{ji2024sustainable}, the $Cl^-$ ions  polarize the water molecules directly bound to the pNIPAM chains, which induces the hydrophobic interactions. Simultaneously, the salt cations ($Na^+$) occupy the interstitial spaces near the dehydrated pNIPAM chains and bind to the carbonyl oxygen atoms of the amide groups. Furthermore, $Na^+$ effectively screens the negative surface charges of the microgel, reducing the Debye length ($\kappa^{-1}$) and leading to the collapse of the electrical double layer that provides colloidal stability. Thermodynamically, a higher external $NaCl$ concentration ($c_s^{out}$) creates a hypertonic environment that minimizes the Donnan osmotic contribution. Because the Donnan effect typically promotes microgel swelling through internal osmotic pressure, its reduction triggers the expulsion of water—a process known as osmotic deswelling—leading to microgel collapse. This mechanism explains the observed shrinkage at low temperatures where the polymer would otherwise remain solvated. These combined factors favor polymer-polymer interactions as $NaCl$ concentration increases, causing collapse of the microgel, and significantly reduce VPTT, as shown in Fig. \ref{fig1}. A comprehensive analysis of the factors affecting the VPTT of pNIPAM microgels, including the network architecture and composition and ionic strength of the medium, is included in this section.

As MBA concentration increases, VPTT monotonically increases in all architectural variants in the absence of added salt, as shown in Fig. \ref{fig1}. For example, ULC microgels [MBA/NIPAM = 0], homogeneously crosslinked (HC) microgels [MBA/NIPAM = 0.020], and the highest crosslinked microgel [MBA/NIPAM = 0.097]  exhibit VPTTs of  32.03$^{\circ}C$, 32.55$^{\circ}C$ and 34.15$^{\circ}C$ respectively. Although the increase is almost 2$^{\circ}C$, the effect of MBA concentration is clearly evident in pure water (0 $mM$ $NaCl$). However, when $NaCl$ is introduced, there is a universal decrease in the VPTT. Specifically, VPTT reduces to 30.13$^{\circ}C$ for ULC microgels, 30.93$^{\circ}C$ for the highest crosslinked (dense) microgel, and 30.71$^{\circ}C$ for HC microgels at 100 $mM$ $NaCl$. For ULC microgels, flocculation occurs at around 32.5$^{\circ}C$ even at 10 $mM$ $NaCl$ concentrations. For other microgels, flocculation occurs at 100 $mM$ concentrations and higher temperatures, influenced by crosslinking density. Microgels with lower crosslinking densities aggregate at around 32$^{\circ}C$, while those with higher crosslinking densities [MBA/NIPAM = 0.077 and 0.097] and HC microgels aggregate at around 34$^{\circ}C$. Data points beyond these points for the onset of flocculation were cropped and fitted with the phenomenological model to obtain VPTT\cite{ks2025revisiting,sennato2021double}. The fitting was performed on the stable, pre-flocculation regime, where DLS accurately captures the microgel particle collapse before inter-particle attractive forces becomes dominant. The VPTT represents the temperature at which a significant size reduction occurs just before flocculation at a given ionic strength, as determined by analyzing the data points prior to the flocculation; this definition is adopted for the VPTT at 100 $mM$ $NaCl$. While pointing out the effect of crosslinking density, soft networks show modest ionic strength sensitivity, while highly crosslinked microgels exhibit amplified sensitivity, resulting in a more significant VPTT reduction. The spatial distribution of crosslinking density appears to have a negligible impact on the sensitivity of the VPTT to ionic strength. For the HC microgel ([MBA/NIPAM] = 0.020), the reduction in VPTT is comparable to that of core-corona microgels with similar average crosslinking densities ([MBA/NIPAM] = 0.007 and 0.031). Furthermore, the absence of a crosslinker has a minimal effect on this sensitivity; the VPTT reduction of the ULC microgels remains comparable with that of microgels containing low concentrations of MBA. Increased ionic strength decreases entropic contributions (favoring swelling), while elastic contributions (favoring retraction of the network and hence deswelling) become significant, leading to a greater VPTT depression in highly crosslinked microgels. Notably, increasing $NaCl$ to 100 $mM$ causes VPTT to drop to a range of 30-30.93 $^{\circ}C$ for all microgel formulations.
 
\subsection{\label{sec:level13}Impact of network architecture and composition on swelling behavior and salt tolerance}

\begin{figure*}[t] 
\centering
  \begin{minipage}{0.32\textwidth}
    \includegraphics[width=\textwidth]{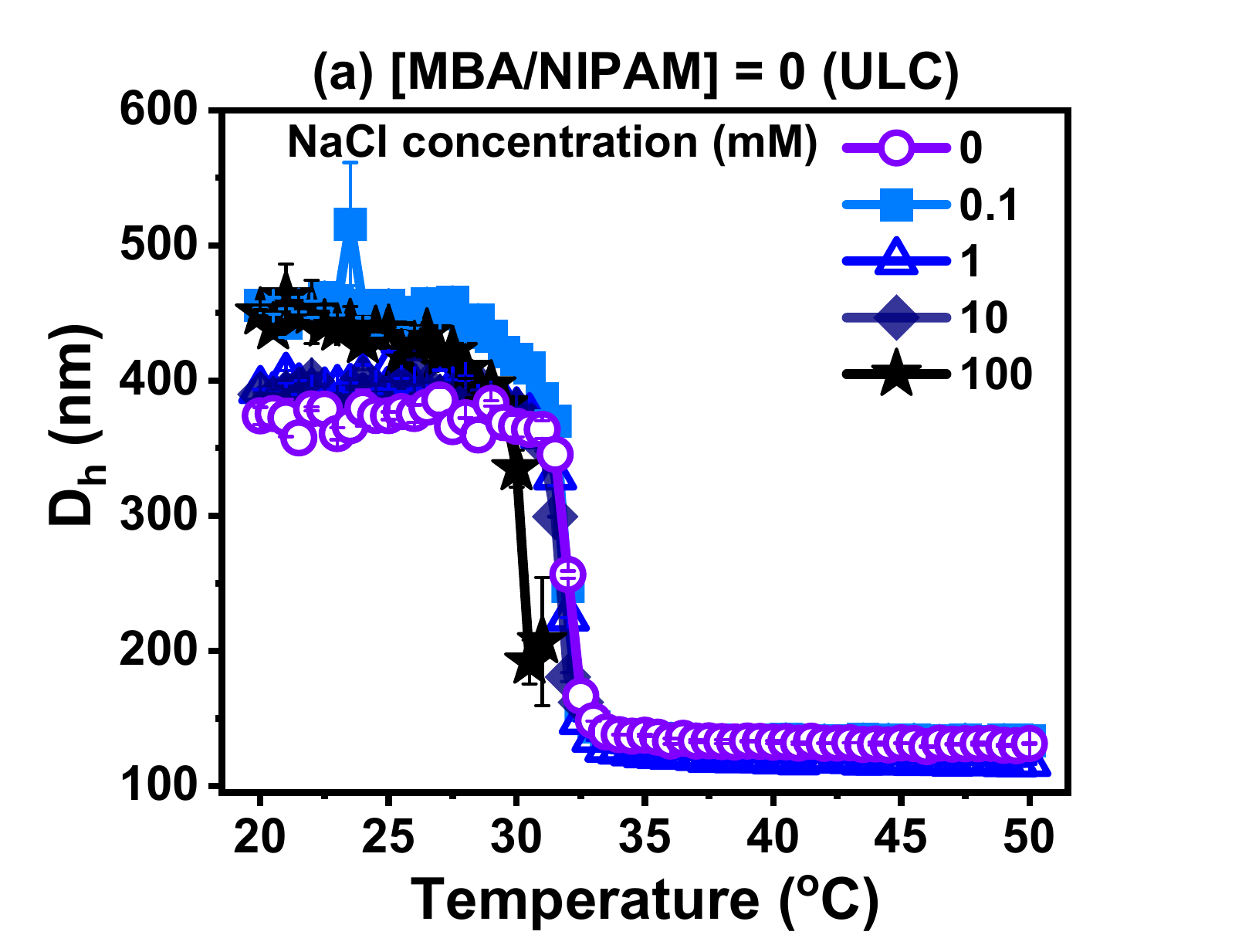}
  \end{minipage}
  \hfill
  \begin{minipage}{0.32\textwidth}
    \includegraphics[width=\textwidth]{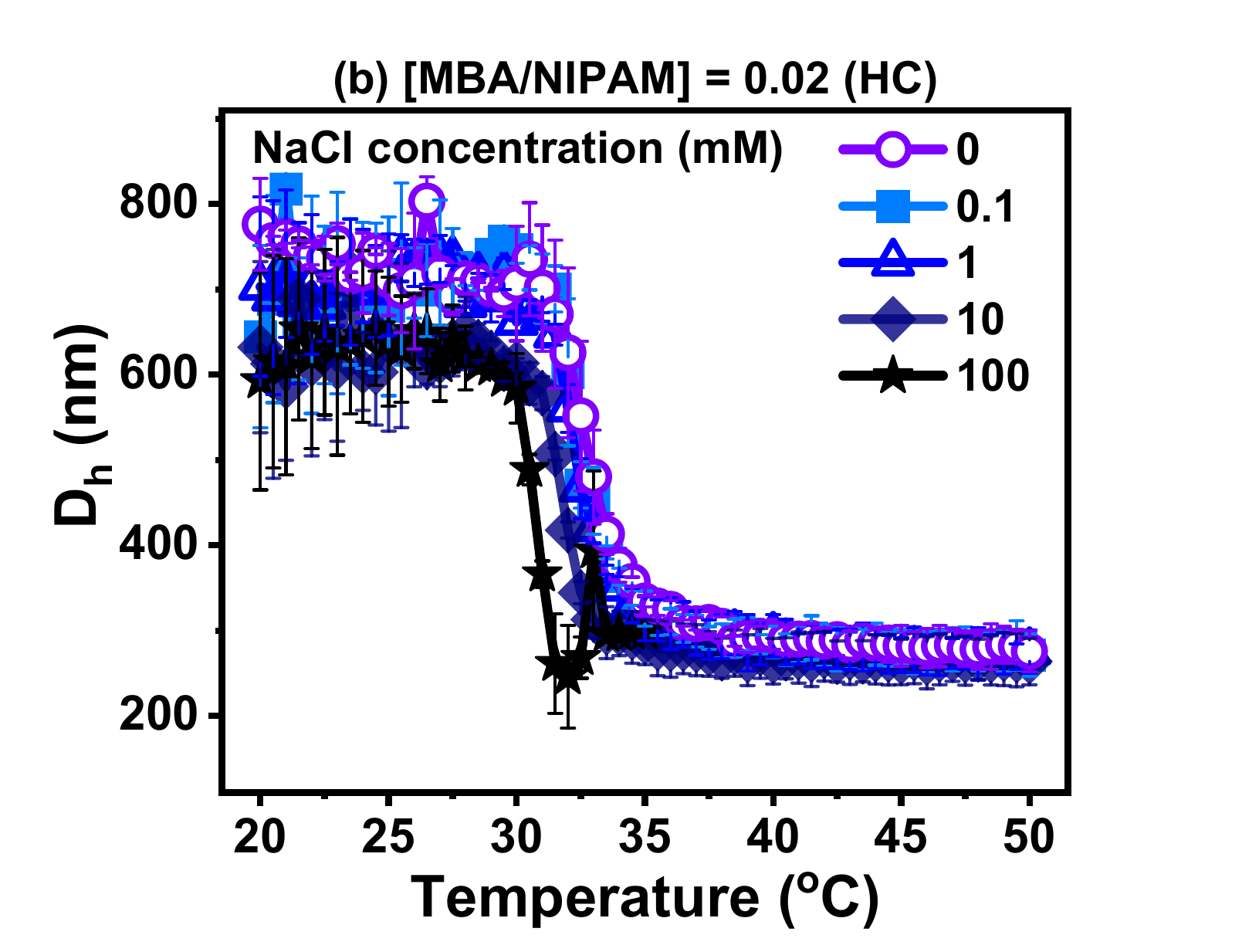}
  \end{minipage}
  \hfill
  \begin{minipage}{0.32\textwidth}
    \includegraphics[width=\textwidth]{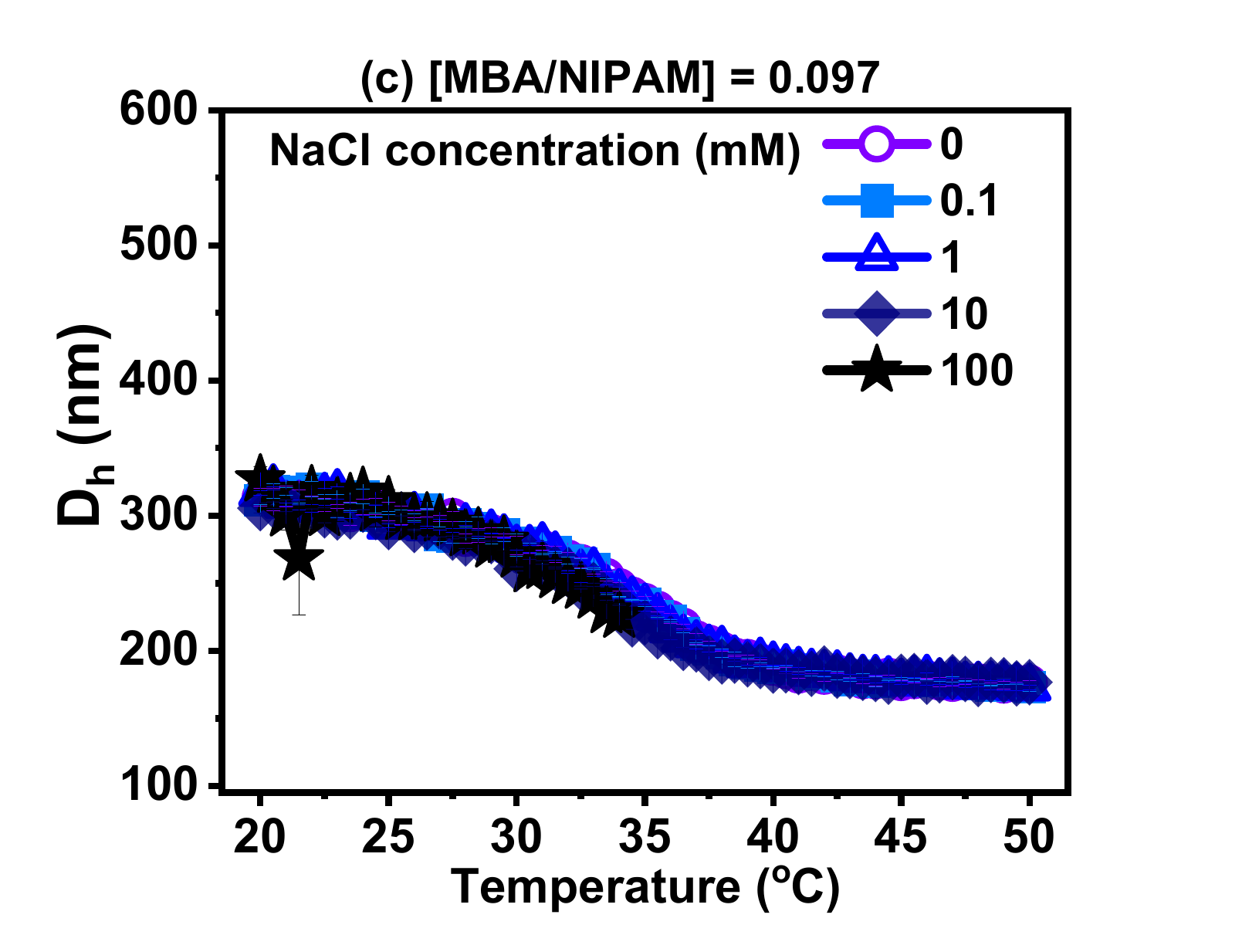}
  \end{minipage}
  
  \vspace{0.2cm} 

  \begin{minipage}{0.32\textwidth}
    \includegraphics[width=\textwidth]{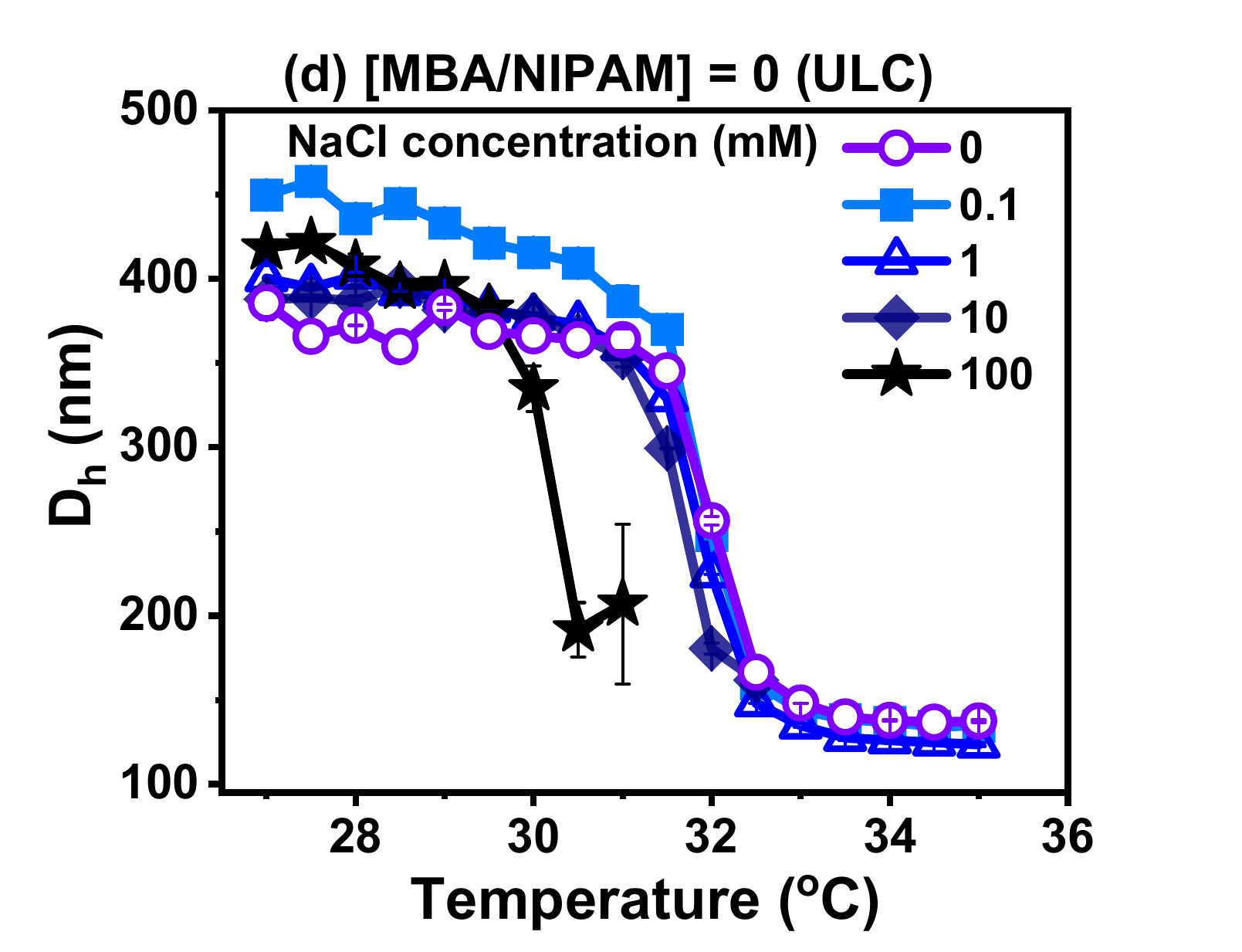}
  \end{minipage}
  \hfill
  \begin{minipage}{0.32\textwidth}
    \includegraphics[width=\textwidth]{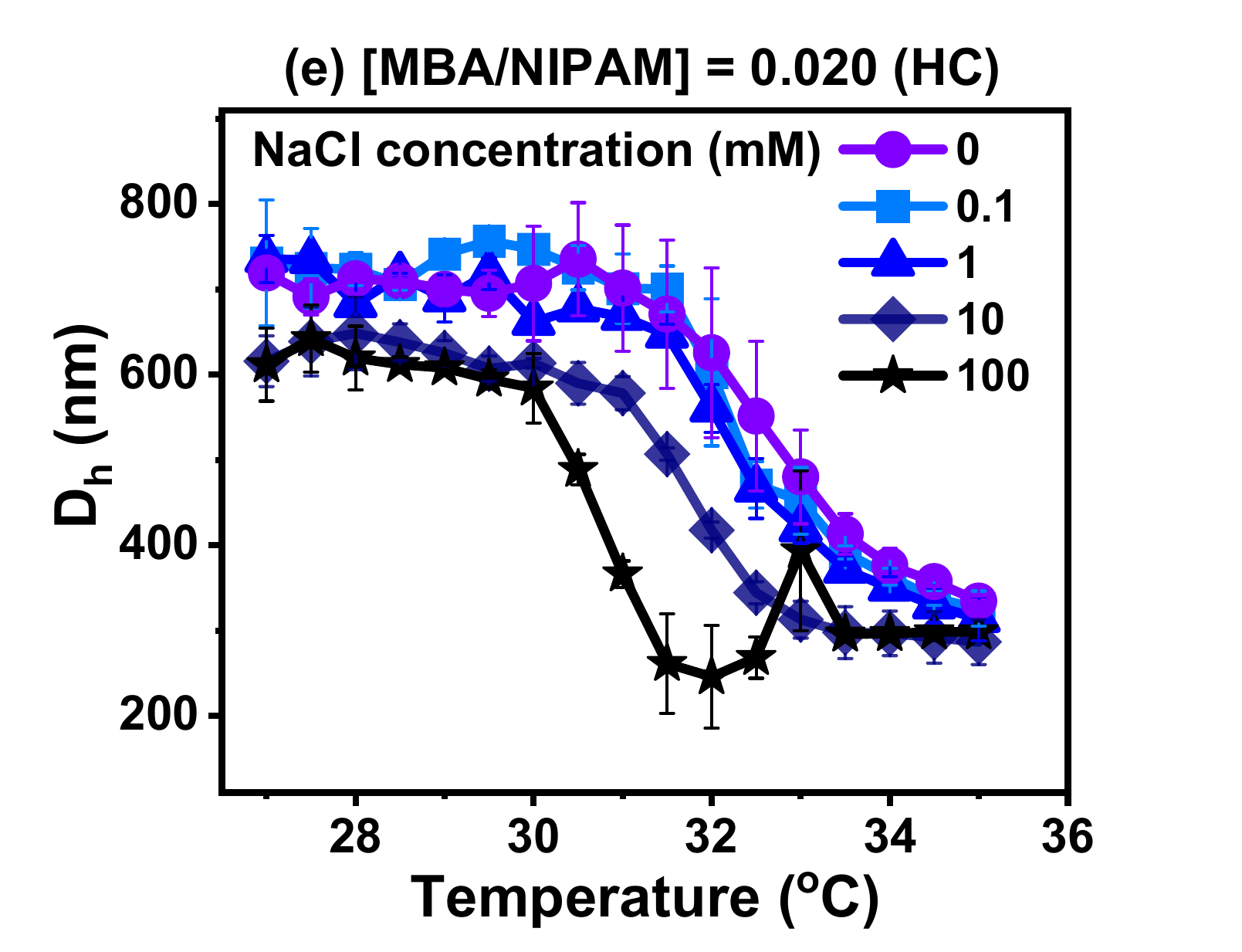}
  \end{minipage}
  \hfill
  \begin{minipage}{0.32\textwidth}
    \includegraphics[width=\textwidth]{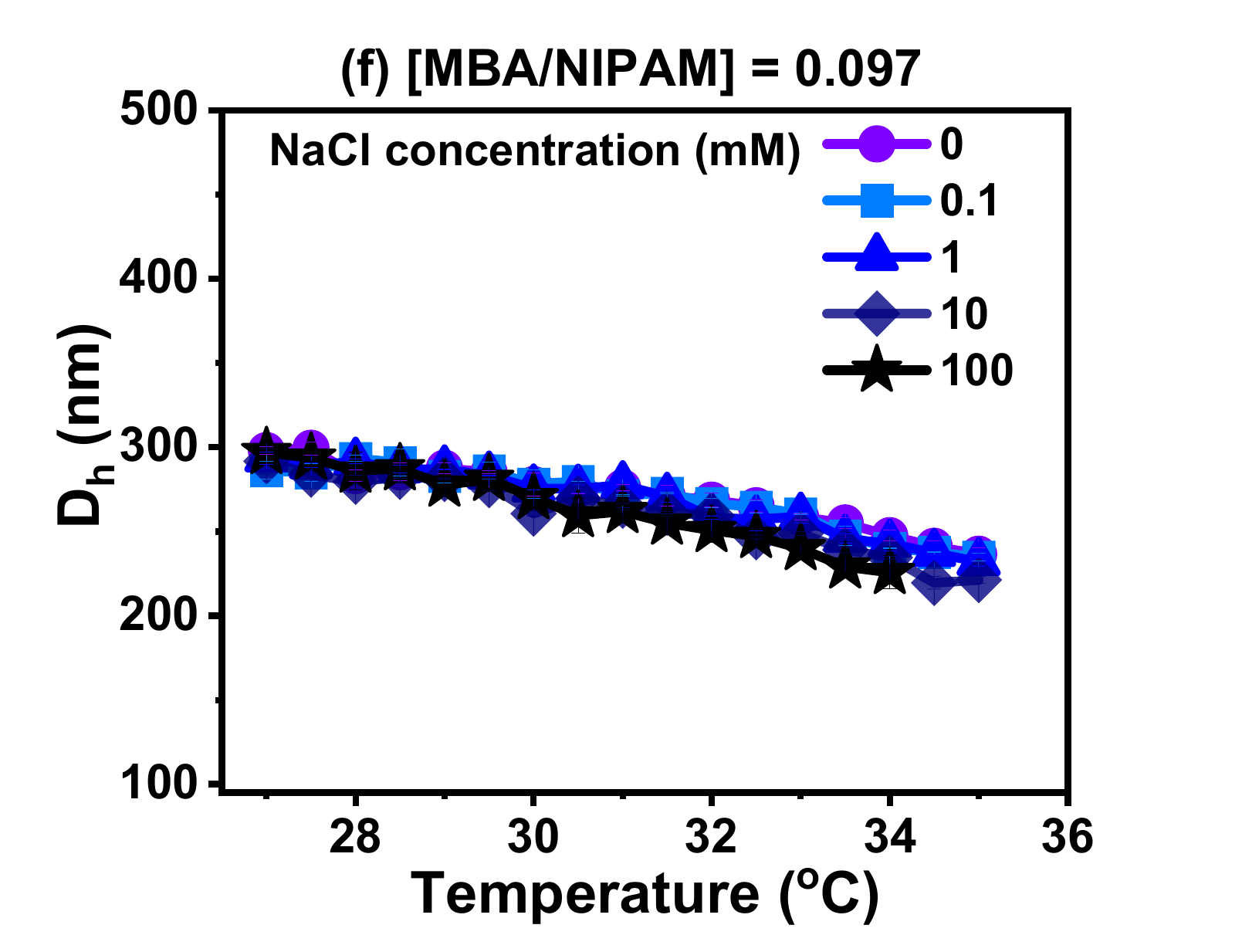}
  \end{minipage}
  
  \caption{Hydrodynamic diameter as a function of temperature for microgels with varying formulations and $NaCl$ concentrations. Top panel: Full temperature range (20–50$^{\circ}C$); Bottom panel: Zoomed-in view near the transition region. (a,d) ULC microgels, (b,e) High-crosslinking density core-corona microgels, (c,f) Homogeneously crosslinked (HC) microgels. Symbols represent different NaCl concentrations: open circle (0 $mM$, pure water, violet), square (0.1 $mM$, light blue), open triangle (1 $mM$,blue), diamond (10 $mM$, navy blue), and star (100 $mM$,black)}
  \label{fig2}
\end{figure*}

Beyond the transition temperature, the functional robustness of microgels is defined by their ability to maintain structural stability against osmotic fluctuations in saline environments. We investigated salt tolerance by examining the absolute size changes and introducing a normalized size metric, defined as, $\lambda_D$ = $D_{h,salt}/D_{h,water}$. This metric quantifies the extent of swelling or deswelling, allowing for a direct comparison of how different microgel formulations maintain their size in saline conditions. Furthermore, the swelling ratio ($\alpha_{\text{size}}$ = $D_{h,swollen}/D_{h,collapsed}$) remains a fundamental property of the microgel, as it quantifies the capacity for thermal volume contraction (the reduction in microgel volume as temperature increases). As reported in our previous work\cite{ks2025revisiting}, this ratio is inversely proportional to the crosslinking density. This trend indicates that higher crosslinking densities restrict the conformational entropy and stretchability of the network, thereby reducing swellability. Consequently, the overall swelling behavior is a synergistic result of the network topology, temperature, and ionic strength.

\subsubsection{Absolute swelling behavior}

Fig. \ref{fig2} presents the temperature-dependent hydrodynamic diameter at varying ionic strengths of the dispersion medium for three representative microgels with different crosslinking densities and crosslinker distributions, spanning the full crosslinking density spectrum investigated in this manuscript.  

ULC microgels (Fig. \ref{fig2} (a, d)) demonstrate pronounced sensitivity to salinity. In the swollen state, ULC microgels exhibit an increase in size with added salt, an anomalous salting-in effect likely facilitated by the negligible elastic restorative force of the loosely connected network. The absence of MBA crosslinks prevents the formation of a strictly affine network, allowing the chains to expand freely. ULC microgels have a size of $\approx$ 373 $nm$ in the swollen state and collapse to $\approx$ 131 $nm$ at 0 $mM$ $NaCl$ concentration. On analyzing the $\alpha_{\text{size}}$, across different $NaCl$ concentrations, it varies in the range of 2.85 to 3.41. Though there is no trend, the $\alpha_{\text{size}}$ increased slightly for ULC when $NaCl$ is introduced into the solvent. As already discussed above, the adsorption of salt ions onto the polymer backbone—specifically $Na^+$ binding to amide oxygens and $Cl^-$ polarizing the hydration shell—initiates a tug-of-war between intra-chain electrostatic repulsions and the network's retracting forces. In ULC microgels, the retracting elastic forces are insufficient to counteract these inter-ionic repulsions and the resulting internal osmotic pressure, leading to the observed expansion at room temperature. However, as solvent quality changes with temperature, approaching VPT, the salting-out effects begin to dominate. This results in more reduction in size compared to the transition in pure water. Notably, in the collapsed state, ULC microgels exhibit deswelling as $NaCl$ concentration increases. As mentioned in the previous section, for ULC microgels, due to the inherent lack of structural constraint, flocculation occurs at around 32.5$^{\circ}C$ even at 10 $mM$ $NaCl$ concentrations; those data were not used for further analysis.

HC microgels (Fig. \ref{fig2} (b, e)) synthesized via surfactant-free semi-batch polymerization; these particles exhibit a significantly larger hydrodynamic diameter ($\approx$ 700 $nm$ in the swollen state and collapse to $\approx$ 275 $nm$, in pure water). This size difference, compared to other microgels synthesized with surfactant, highlights the critical role of surfactants in determining particle size during synthesis\cite{hu2011synthesis,wedel2017role}. The range of $\alpha_{\text{size}}$ for HC, across different $NaCl$ concentrations, is 2.81 to 2.37 and the $\alpha_{\text{size}}$ slightly decreased when $NaCl$ is introduced. Despite their size, HC microgels demonstrate a pronounced salting-out effect across all temperature regimes. This high sensitivity is possibly attributed to the homogeneous crosslinker distribution, which ensures that the entire microgel volume responds uniformly to the external osmotic pressure gradient.

In contrast, highly crosslinked core-corona microgels (Fig. \ref{fig2} (c, f)) exhibit significant salt tolerance, maintaining a relatively stable hydrodynamic size even at high ionic strengths. These microgels have a size of $\approx$ 313 $nM$ in the swollen state and collapse to $\approx$ 176 $nM$ in pure water. The $\alpha_{\text{size}}$ change is considerably less (ranges from 1.72 to 1.83) for the  highly crosslinked microgels with $NaCl$ concentration. The high density of crosslinks creates a stiff elastic network that effectively resists the osmotic pressure changes that would otherwise induce drastic swelling or deswelling. This structural rigidity acts as a mechanical constraint, limiting the conformational changes of the polymer chains in response to ionic variations. Consequently, the core-corona architecture provides a functional robustness, ensuring structural stability in saline environments where ULC and HC microgels undergo significant hydrodynamic volume changes.

The temperature-dependent hydrodynamic diameter at varying ionic strength for other microgel formulations, including other crosslinking density variants (Fig. \ref{SI:007} to Fig. \ref{SI:0.077}), are provided in the supplementary material. This comparison demonstrates that the sensitivity of a microgel to its environment is governed by network topology. While softer networks are highly responsive to changes in ionic strength due to their low elastic restorative force, dense or heterogeneously crosslinked (core-corona) networks provide structural robustness that effectively shields the microgel from drastic volumetric transitions.

\subsubsection{Quantifying microgel salt tolerance ($\lambda_D$)}
\begin{figure}[h!]
\centering
  \includegraphics[height=6cm]{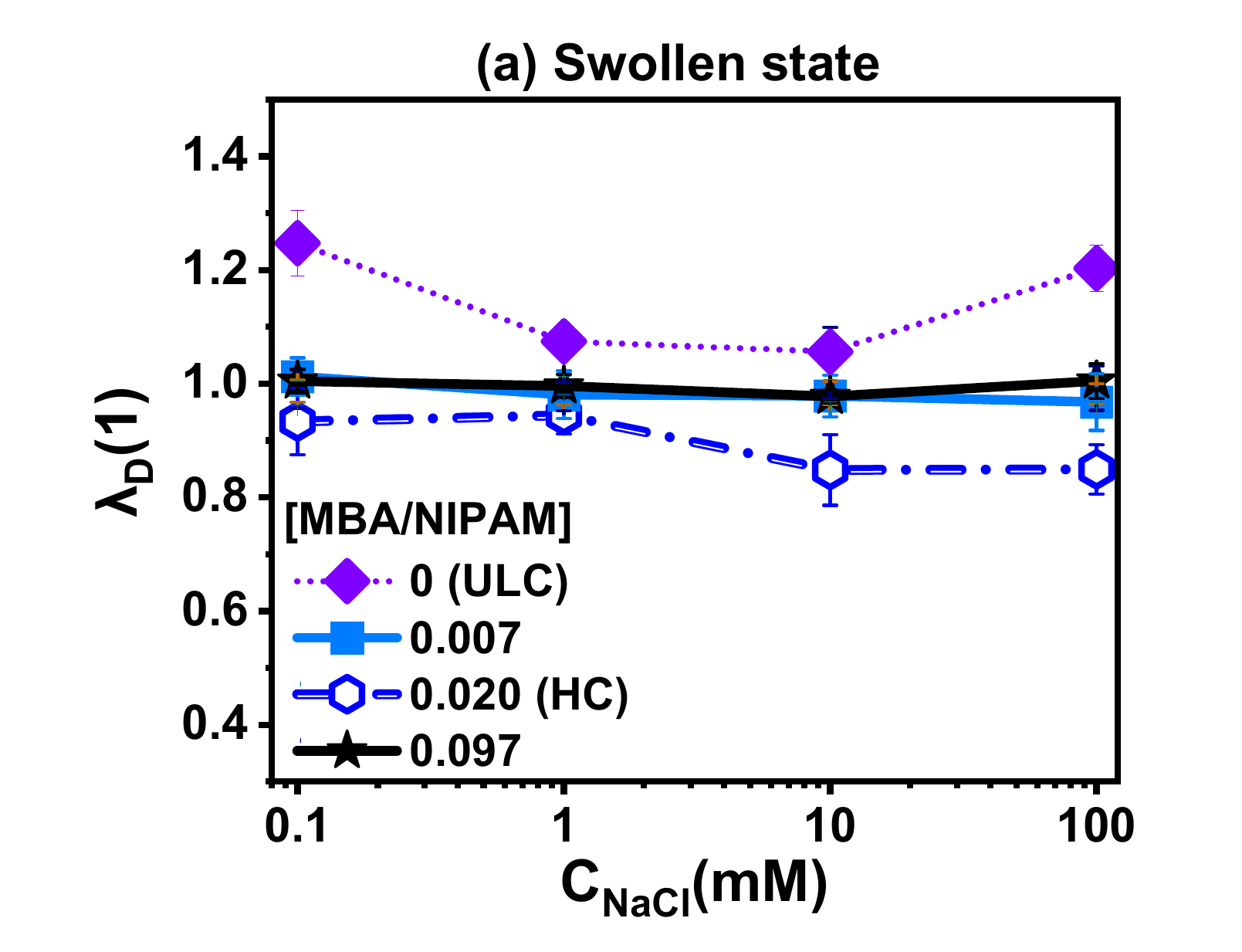}
   \includegraphics[height=6cm]{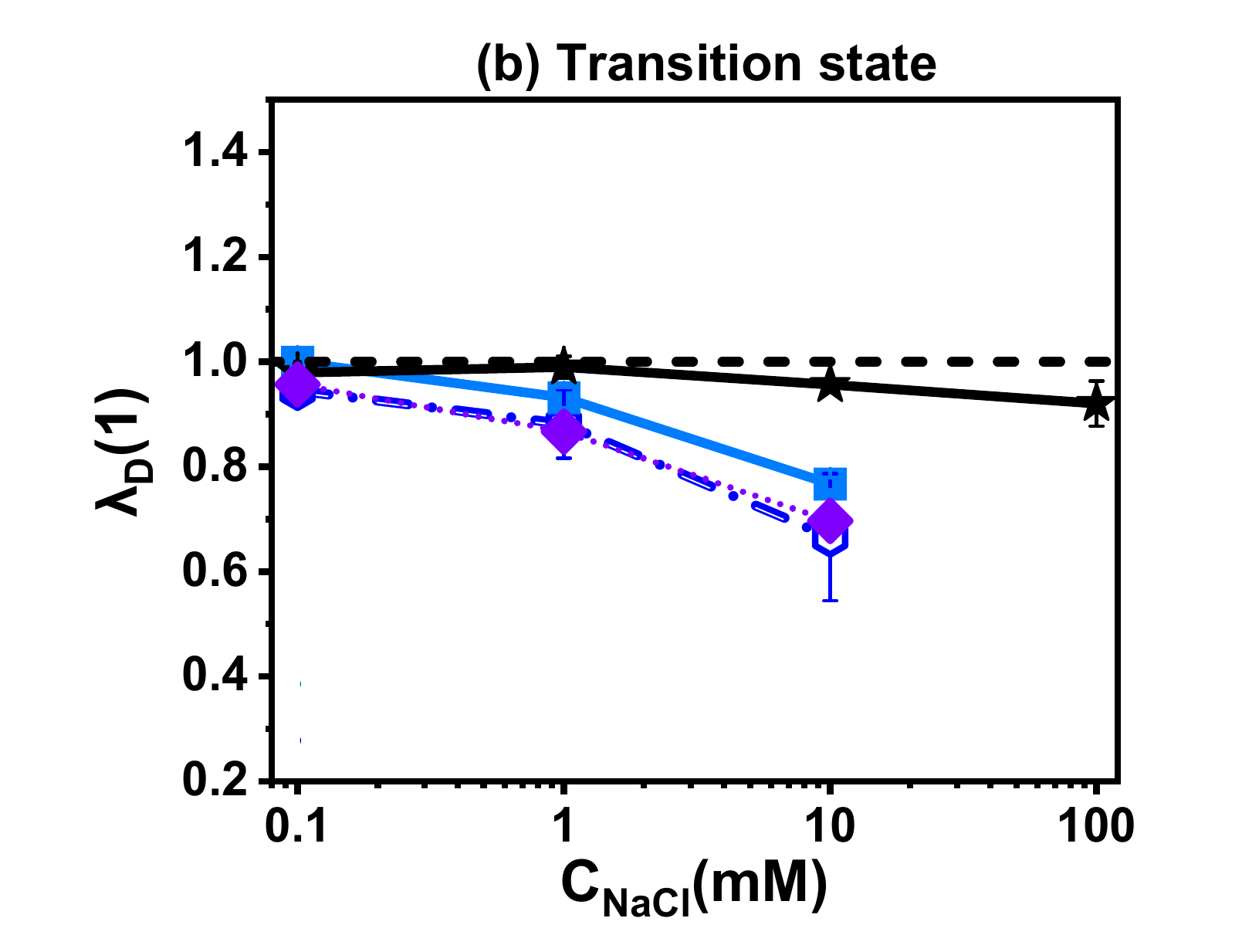}
   \includegraphics[height=6cm]{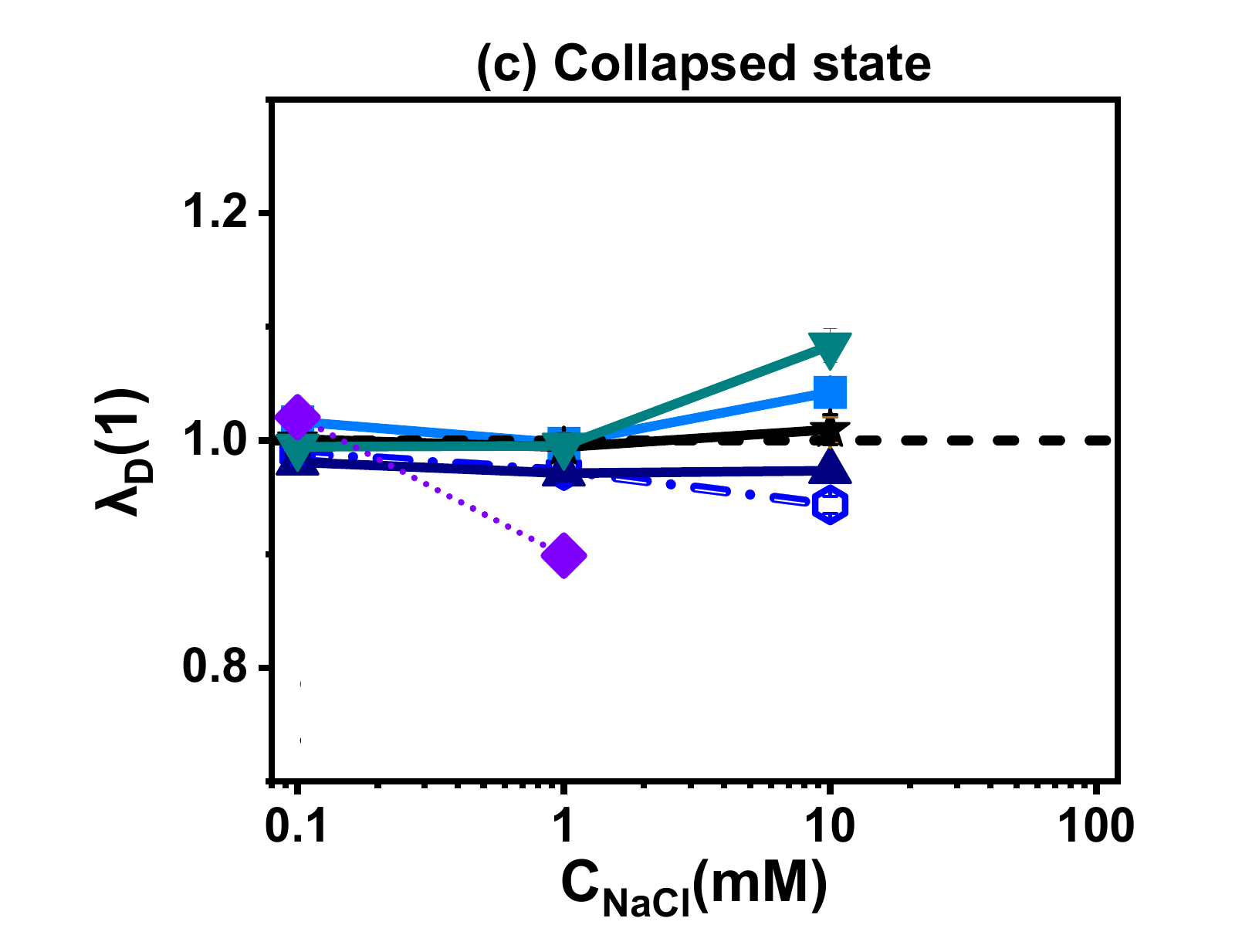}
  \caption{Normalized Size ($\Lambda$) as a metric for salt tolerance across three thermodynamic states: (a) swollen state, (b) transition state, and (c) collapsed state. $\Lambda$ = 1.0 indicates the ideal case, the highest salt resistance; $\Lambda$ $<$ 1.0 indicates salt-induced deswelling. Soft architectures (ULC, homogeneous) show significant osmotic deswelling, while rigid core-corona networks maintain structural integrity. Symbols: open square (homogeneous crosslinking-HC microgel), triangle (low initiator concentration), inverted triangle (high initiator concentration), open circle with dotted line (ULC microgel), closed dark yellow circle (highest crosslinking densidensity), and green circle (lowest crosslinking density).}
  \label{fig3}
\end{figure}

To provide a rigorous metric for observations on the influence of ionic strength on microgel salt tolerance, we quantified salt tolerance using the parameter $\lambda_D$, defined as the normalized ratio of microgel size in salt solution to its size in pure water.  Fig. \ref{fig3} quantifies $\lambda_D$ for microgels in three distinct thermodynamic states, specifically  the swollen (Fig. \ref{fig3} (a)), VPT (Fig. \ref{fig3} (b)), and collapsed (Fig. \ref{fig3} (c)) states at varying ionic strengths. Figure \ref{fig3}, shows that when $\lambda_D$ is 1, it is an ideal case where there is no change in the microgel size with ionic strength of the medium due to salting-out, salting-in, or flocculation, implying a high salt tolerance. 

In the swollen state (Fig. \ref{fig3} (a)), analysis was performed below the transition temperature (20-25$^{\circ}C$); the microgels exhibit highly architecture-dependent responses. The ULC microgel variant demonstrates significant salting-in effects; specifically, the ULC microgel reaches a $\lambda_D$ of $1.25 \pm 0.06$ at 0.1 mM NaCl. This is likely due to an initial difference in osmotic pressure within its loosely bound, self-crosslinked network. Conversely, the HC microgels are the most sensitive to the salting-out effect, with $\lambda_D$ dropping to $0.85 \pm 0.04$ at 100 $mM$ $NaCl$. The highly crosslinked variants maintain robust salt tolerance, with $\lambda_D$ remaining near unity in the swollen state.

The salt sensitivity of the microgel network is maximized during the VPT state (Fig. \ref{fig3}(b)), where analyses were performed at the $VPTT$ of the respective microgels in pure water. Here, the soft architectures like ULC and HC undergo dramatic salt-induced deswelling, with $\lambda_D$ falling to $0.70 \pm 0.01$ and $0.67 \pm 0.12$ at 10 $mM$ $NaCl$, respectively. This suggests that in the transition region, even moderate ionic strength can significantly accelerate the collapse of the microgel. In contrast, the highly crosslinked microgel remains almost unaffected ($\lambda_D = 0.96 \pm 0.001$), demonstrating that a rigid network provides a robust elastic backbone that resists salt-induced volume changes even at the VPT state. Furthermore, only the highly crosslinked microgel withstands flocculation at 100 $mM$ $NaCl$ in the VPT state, further confirming its superior structural stability and lower salt sensitivity. 

In the collapsed state, the analysis was performed above the transition temperature ((45-50$^{\circ}C$) (Fig. \ref{fig3} (c)); all microgel formulations exhibited significant salt tolerance. Regardless of the microgel formulation, $\lambda_D$ remains between $0.90$ and $1.08$ across the tested concentration range. This suggests that, the high polymer density in the collapsed state effectively shields the microgels from further osmotic changes. Notably, while the ULC microgels still exhibit the lowest $\lambda_D$ values in the collapsed state, the deviation is minimal compared to the swollen and VPT states, indicating that structural differences are significantly minimized once the polymer chains reach a fully collapsed configuration.

The ULC microgels are the most sensitive; the lack of MBA crosslinks results in a negligible elastic restorative force, allowing for both anomalous swelling and then complete collapse. Conversely, the HC microgels are characterized by their homogeneous crosslinking density; this makes the entire volume susceptible to osmotic deswelling and significant salting-out effects. The core-corona microgels emerge as the robust middle ground. By concentrating MBA crosslinks in the center, this architecture establishes a dense core that prevents the particle from collapsing beyond a critical threshold. This design allows the corona to maintain functional interactions with the solvent while the stiff core preserves the microgel's overall structural integrity against osmotic fluctuations.


\subsection{\label{sec:level141}Reversibility and thermal hysteresis}

\begin{figure}[h!]
\centering
 \includegraphics[height=6cm]{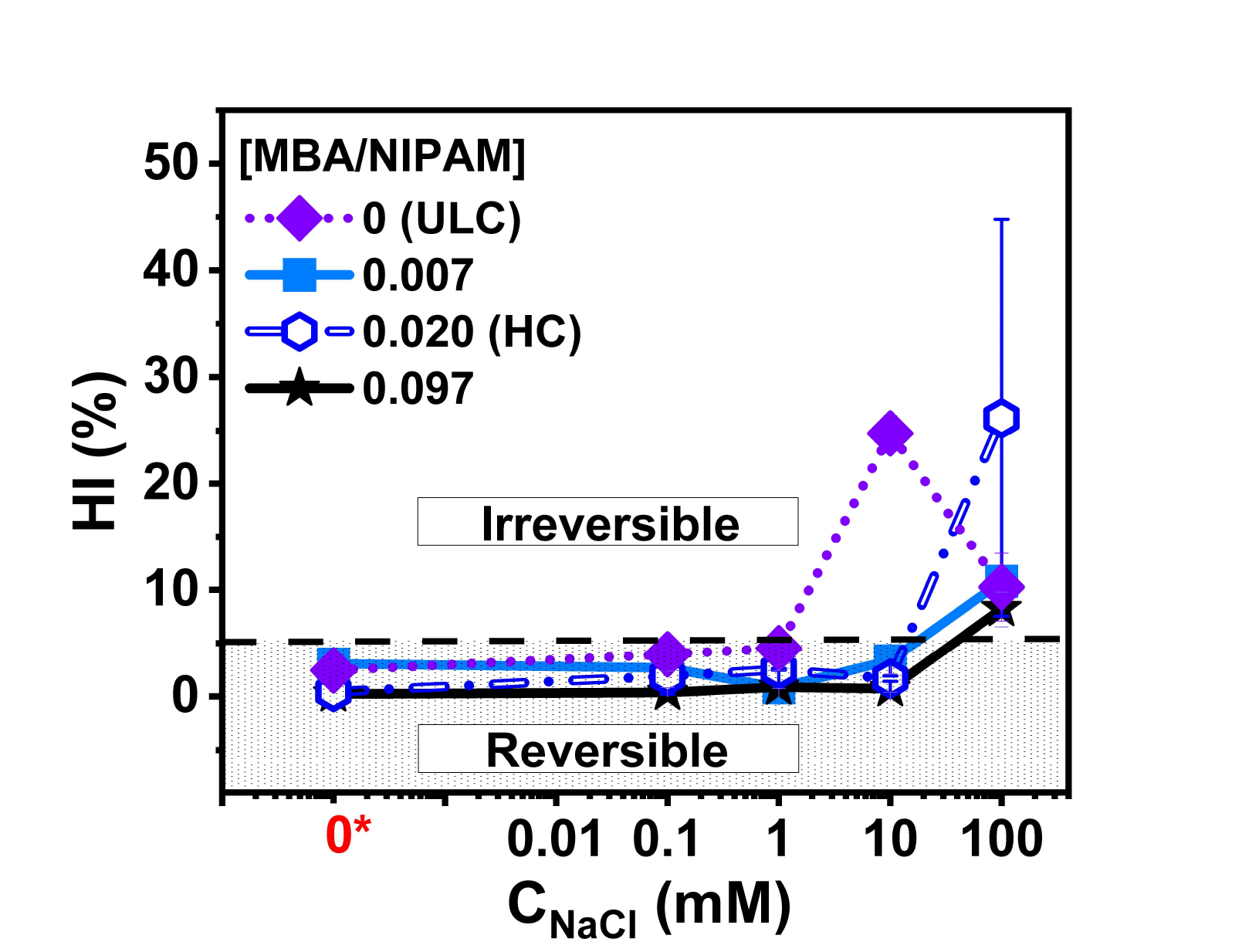}
\caption{The Hysteresis Index (HI), derived from heating–cooling cycles across varying ionic strengths ($NaCl$ concentration), is significantly influenced by both the MBA crosslinker concentration and its specific spatial distribution within the microgel (ULC, homogeneous, core-corona). The x-axis is logarithmic, with "0*" plotted as a nominal notation for pure water (0 $mM$ $NaCl$). HI = 10\% can be considered as the the operational threshold between the reversible regime (HI $<$ 5\%) and the irreversible aggregation regime (HI $>$ 5\%). Symbols: closed violet diamonds (ULC), closed blue squares ([MBA/NIPAM] = 0.007), open blue circles (HC), and black stars ([MBA/NIPAM] = 0.097).}
\label{fig4}
\end{figure}

The pNIPAM microgels showed good thermoreversibility in pure water \cite{ks2025revisiting}. These colloidal microgel suspensions possibly exhibit thermal hysteresis in saline environments due to irreversible flocculation.  To quantify this behavior, we use a metric, the hysteresis index (HI), where a low HI indicates perfect recovery (the cooling curve overlaps the heating curve) and a high HI shows hysteresis or irreversible flocculation in saline medium. Fig. \ref{fig4} presents the  HI against $NaCl$ concentrations on a semi-log scale, establishing a boundary between reversible and irreversible regimes at a 5\% HI cutoff. This threshold represents the transition from a thermodynamically reversible phase transition to a kinetically trapped flocculated state. 

The stability of the microgel is a balance between electrostatic repulsion and hydrophobic interactions in saline media. In the low ionic strength regime, it is found that all microgels—regardless of architecture or crosslinking density—exhibit robust reversibility with HI values consistently below 5\%. At low salt concentrations, increasing temperature makes water a poor solvent for the microgel, leading to a dominance of hydrophobic interactions and thus the volumetric collapse of the microgel. Despite collapse, minimal flocculation is observed because the electrostatic repulsion between particles dominates, keeping collapsed microgels well dispersed via electrostatic stabilization. The addition of $NaCl$ induces polarization of the water molecules directly bound to the pNIPAM chains and facilitates specific counter-ion adsorption \cite{ji2024sustainable}. This interaction is sufficiently robust that the resulting flocculation becomes essentially irreversible

As mentioned in the previous sections, flocculation pushed particle sizes beyond the DLS measurable window, increasing the measurement errors. Consequently, these flocculation data points were cropped from the analysis. Fig. \ref{fig4} shows the impact of density and distribution of crosslinking on reversibility. Most of the microgels, except ULC, show a high HI value only at 100 $mM$ $NaCl$, for ULC, even at 10 $mM$ $NaCl$, it shows significant hysteresis. It shows that the microgel with MBA crosslinker and core-corona microstructure behaves almost the same in the aspect of reversible thermoresponsiveness in the ionic media. The presence of a rigid, highly crosslinked core in the core-corona architecture appears to provide a mechanical template that facilitates reversible thermoresponsiveness, a feature notably absent in the ULC and HC variants. The elevated HI observed in ULC microgels likely stems from the absence of an affine crosslinked network, which allows the chains to entangle more easily. In the case of HC microgels, the high HI is attributed to the low and homogeneous distribution of MBA and potentially the absence of a highly crosslinked core (unlike core-corona microgels). This maximizes irreversible flocculation at high salt concentrations. Conversely, core-corona microgels possess a form of elastic memory; the dense core stores elastic energy during volume contraction, providing a restorative force that minimizes flocculation and ensures a low HI. Ultimately, the thermoreversibility of microgels in saline media is significantly influenced by the crosslinking density, the spatial distribution of crosslinks, and the presence of a crosslinker.

\subsection{\label{sec:level15}Flocculation kinetics under extreme ionic stress}

\begin{figure*}[t] 
\centering
  \begin{minipage}{0.45\textwidth}
    \includegraphics[width=\textwidth]{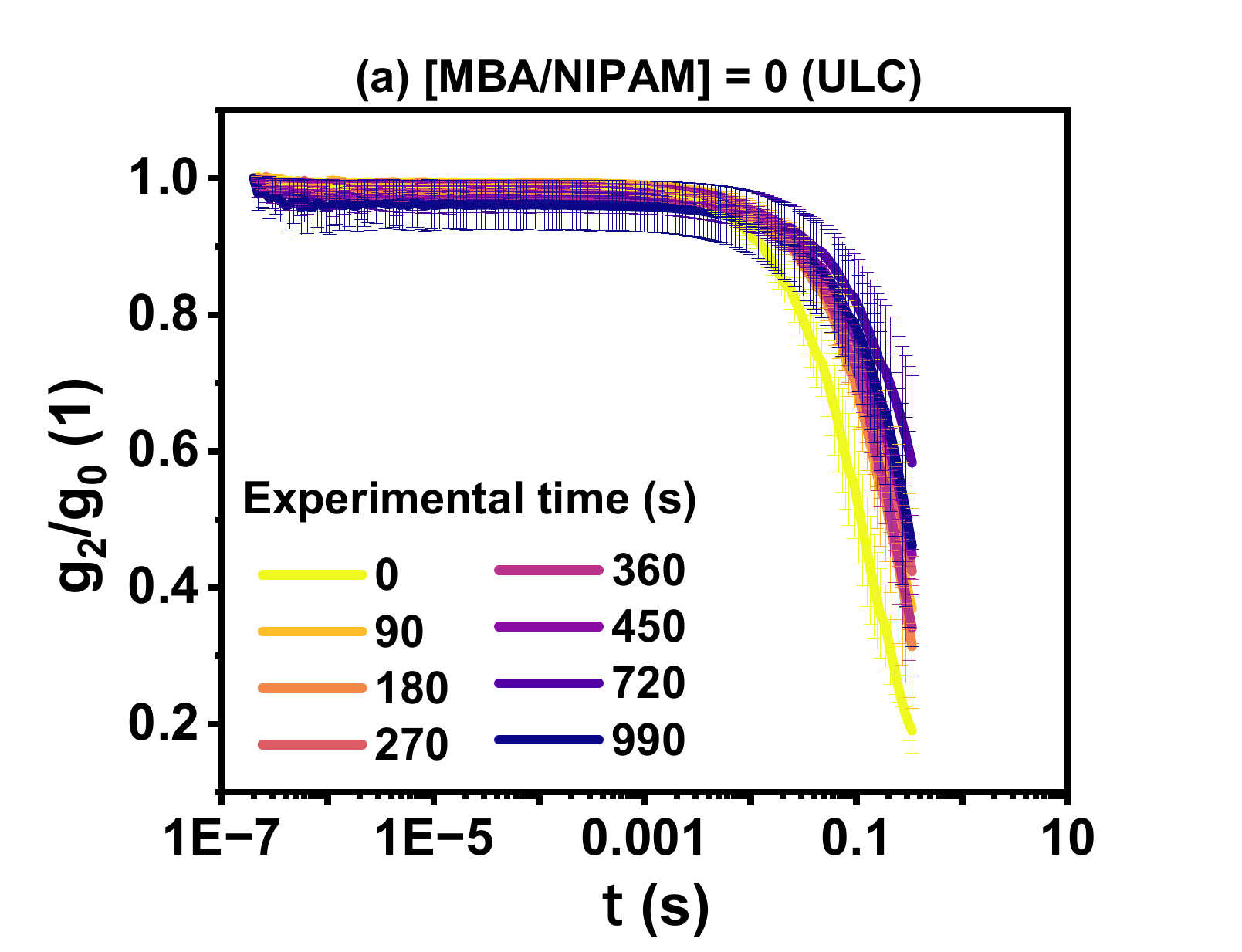}
  \end{minipage}
  \hfill
  \begin{minipage}{0.45\textwidth}
    \includegraphics[width=\textwidth]{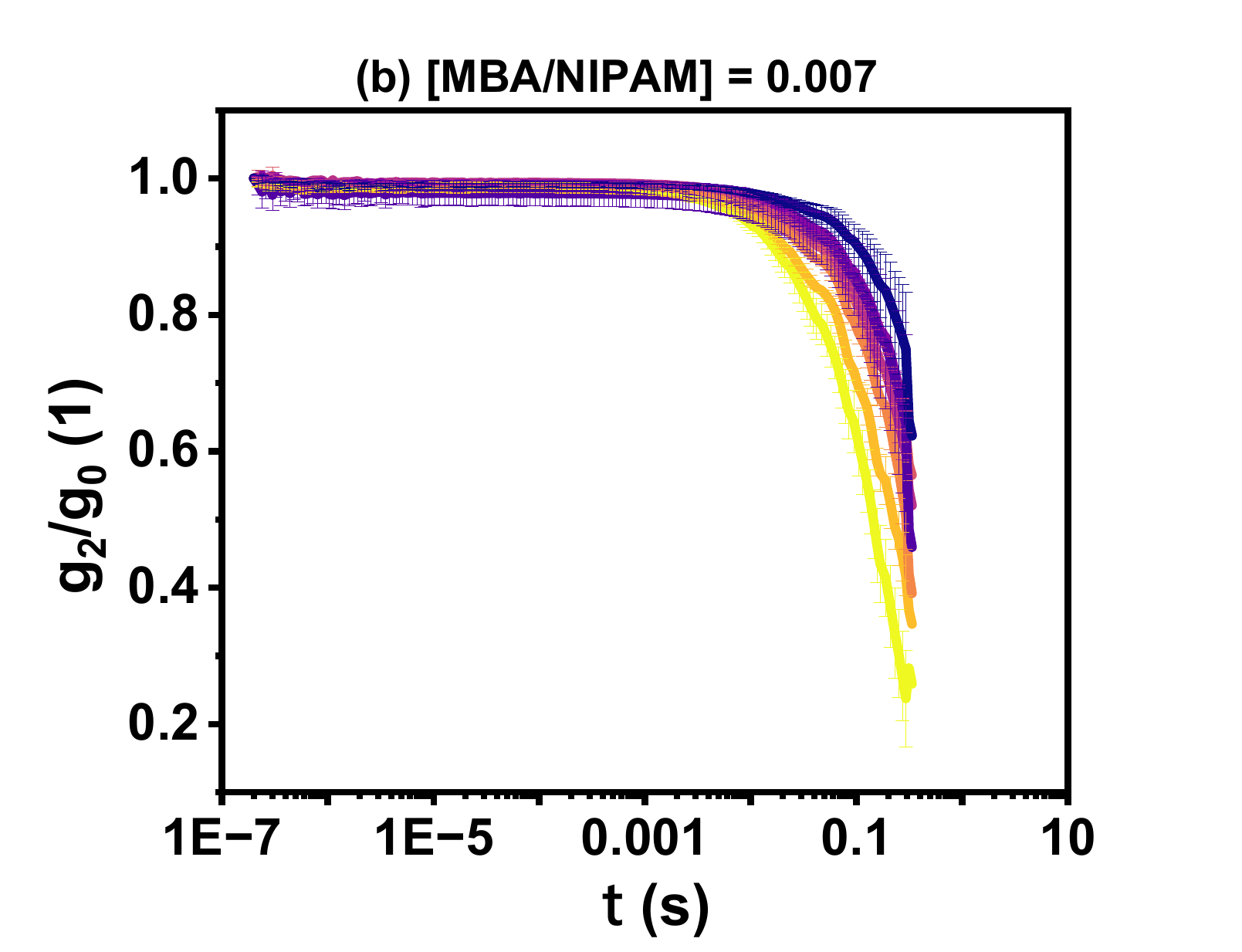}
  \end{minipage}

  \vspace{0.3cm} 

  \begin{minipage}{0.45\textwidth}
    \includegraphics[width=\textwidth]{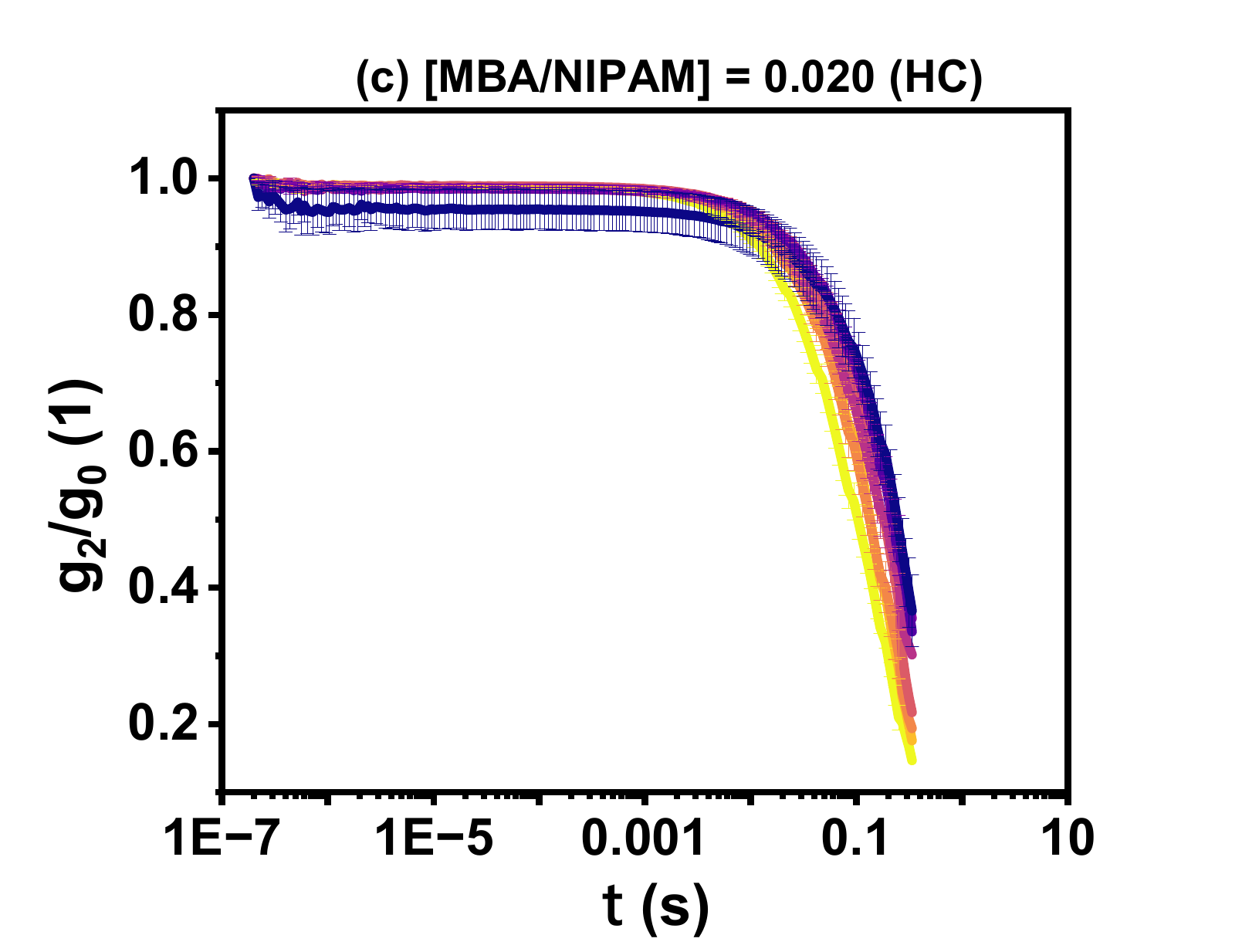}
  \end{minipage}
  \hfill
  \begin{minipage}{0.45\textwidth}
    \includegraphics[width=\textwidth]{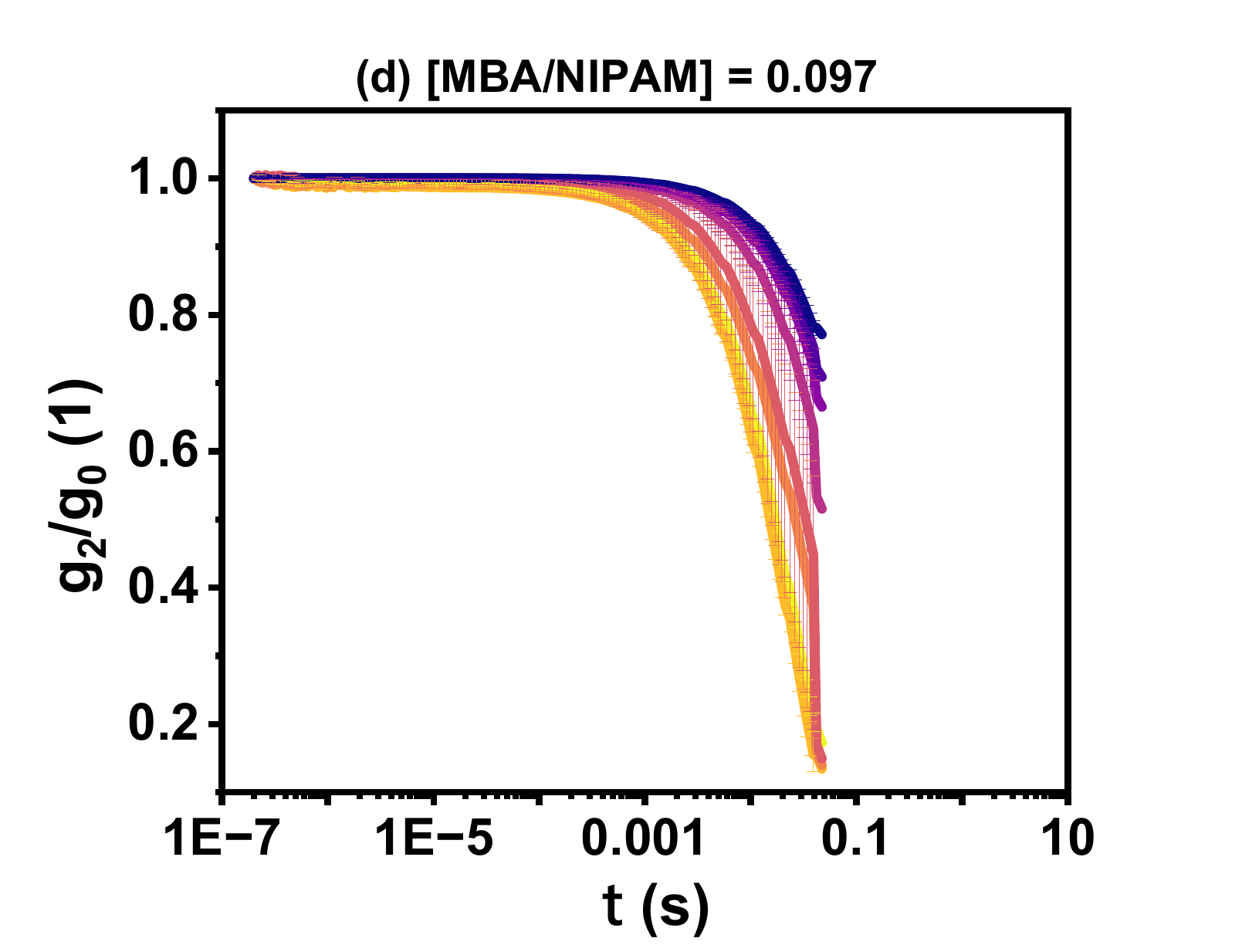}
  \end{minipage}

  \caption{Lag time ($t$) versus normalized autocorrelation function ($g_{2}/g_{0}$)  for different microgel formulations at 1000 $mM$ $NaCl$ at 25°C. Panels a--d represent microgels with varying crosslinking densities and distributions: (a) ULC, (b) [MBA/NIPAM] = 0.007, (c) HC [MBA/NIPAM] = 0.020, and (d) [MBA/NIPAM] = 0.097. Data represent experimental times from 0 to 990~seconds, with colors transitioning from yellow to dark blue}
  \label{fig5}
\end{figure*}

\begin{figure}[h!]
\centering
  \includegraphics[height=6cm]{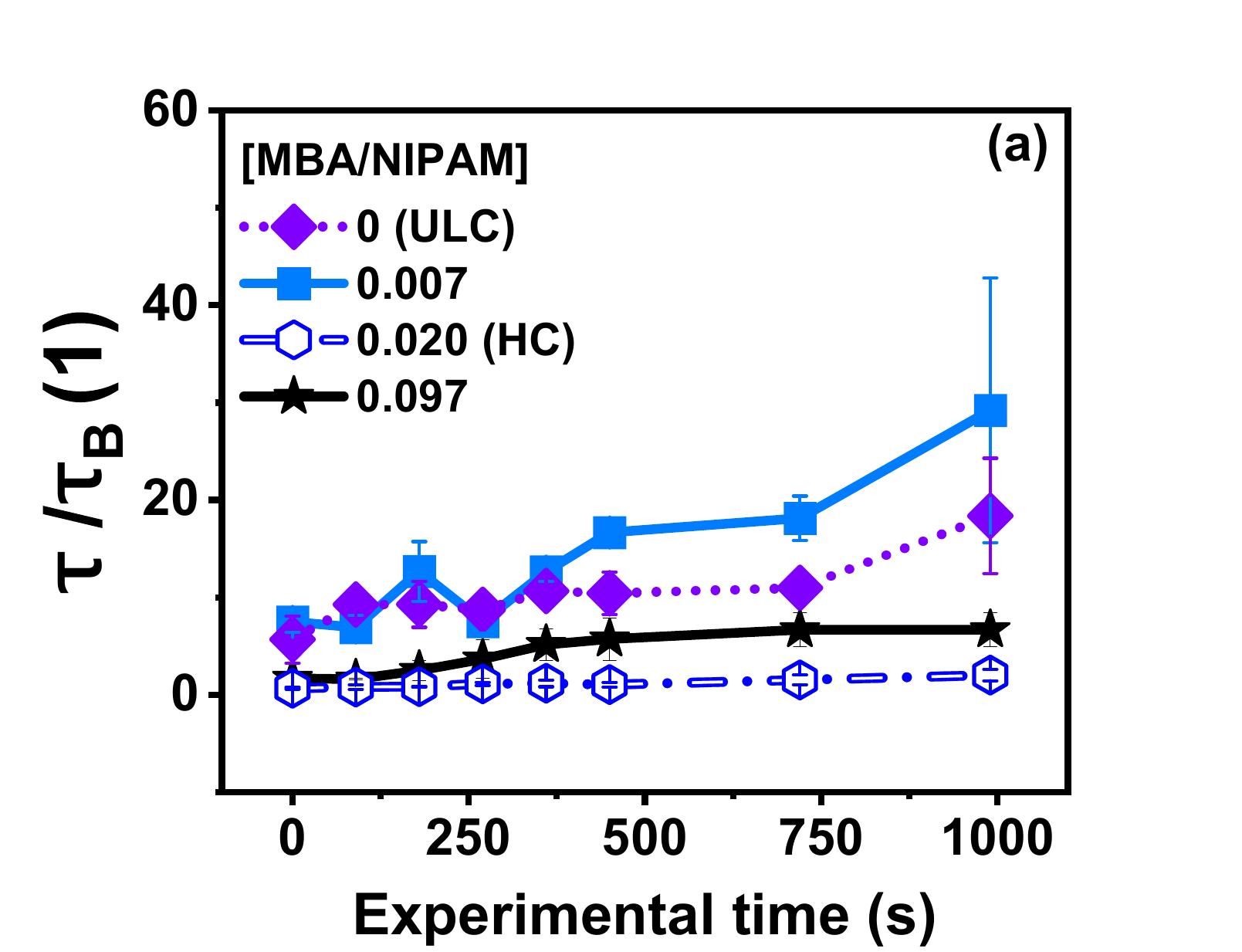}
  \caption{Normalized decay time ($\tau/\tau_B$) versus flocculation time for all microgel architectures under extreme ionic stress (1000 $mM$ $NaCl$). Rapid increases in $\tau/\tau_B$ indicate faster aggregation kinetics, revealing the influence of crosslinker (MBA) concentration and crosslinking distribution (ULC, homogeneous, core-corona). Symbols: closed violet diamonds (ULC), closed blue squares ([MBA/NIPAM] = 0.007), open blue circles (HC), and black stars ([MBA/NIPAM] = 0.097)}
  \label{fig6}
\end{figure}

Flocculation kinetics analysis was performed to quantify the rate of microgel flocculation in extreme saline conditions over a short time interval. At extreme ionic strength (1000 $mM$ $NaCl$), the microgels undergo rapid flocculation at room temperature; analysis at higher temperatures was avoided due to limitations in DLS detection caused by rapid particle growth. By maintaining a constant room temperature, we isolated the effect of salt on solvent quality, preventing temperature-induced phase transitions from interfering with the analysis. We monitored the flocculation behavior of microgels at 1000 $mM$ $NaCl$, using DLS at a forward scattering angle (15$^{\circ}$) and at room temperature (25$^{\circ}C$) for 990 seconds. Upon the addition of 1000 $mM$ $NaCl$ to a 0.33 $mg/mL$ pNIPAM microgel dispersion, flocculation was triggered and monitored via the autocorrelation function $g^{(2)}(t)$. This allowed for the estimation of the characteristic decay time ($\tau$). The high salinity induced immediate flocculation, resulting in a progressive increase in the effective flocculate size. This growth is reflected in the evolution of the normalized intensity autocorrelation functions as a function of lag time ($t$), as shown for the distinct microgel formulations in Fig. \ref{fig5}. As experimental time increases from 0 to 990 seconds, the $g_{2}/g_{0}$ decay slower, shifting the curve to the right towards longer correlation times (Fig. \ref{fig5}). This indicates slowed particle diffusion due to microgel flocculation. Fig. \ref{fig5} displays these $g_{2}/g_{0}$ curves at discrete intervals: 0, 90, 180, 270, 360, 450, 720, and 990 seconds. The panels in Fig. \ref{fig5} reveal distinct trends based on microgel architecture: (a) ULC [MBA/NIPAM] = 0, (b) low crosslinked ([MBA/NIPAM] = 0.007), (c) HC ([MBA/NIPAM] = 0.020), and (d) high crosslinked ([MBA/NIPAM] = 0.097). To quantify the effects of network architecture and initiator concentration on flocculation kinetics, the decay times ($\tau$) were extracted from the autocorrelation functions using the exponential model, as mentioned in the materials and methods section. These values were used to plot normalized decay time ($\tau/\tau_B$) versus experimental time in Fig. \ref{fig6}. To ensure a fair comparison, $\tau$ is normalized by the Brownian time $\tau_B$, accounting for differences in microgel sizes. To quantitatively compare the flocculation rates across different architectures, the slopes ($d(\tau/\tau_B)/dt$) were extracted from linear fits of the experimental data. A higher slope means the effective particle size increases faster, which indicates quicker flocculation.

Fig. \ref{fig6} reveals the impact of network topology on the flocculation kinetics. The microgels with low or non-uniform crosslinking densities exhibited the highest instability at 1000 $mM$ $NaCl$. Specifically, the low-crosslinked variant ([MBA/NIPAM] = 0.007) displayed the most aggressive aggregation with a slope of 0.02 $s^{-1}$, and for the ULC microgels exhibited a slope of 0.01 $s^{-1}$). In these samples, the high standard deviation observed at higher experimental times suggests the formation of large, polydisperse flocculates.  Conversely, the HC microgels demonstrated remarkable colloidal stability, characterized by a nearly horizontal trend and a minimal slope of 0.0013 $s^{-1}$. While increasing the crosslinker concentration to [MBA/NIPAM] = 0.097 resulted in higher stability than the low-crosslinked samples (slope of 0.0059 $s^{-1}$), but it did not reach the near-perfect stability of the HC architecture. The minimal slope values are attributed to the slower dynamics of the HC microgels, a direct consequence of their larger size relative to other variants. Interestingly, microgels that are self-cross linked or have low MBA crosslinker consistently exhibited faster flocculation kinetics.

\subsection{\label{sec:level2}Fits to the experimental data}

\begin{figure*}[t] 
\centering
  \begin{minipage}{0.32\textwidth}
    \includegraphics[width=\textwidth]{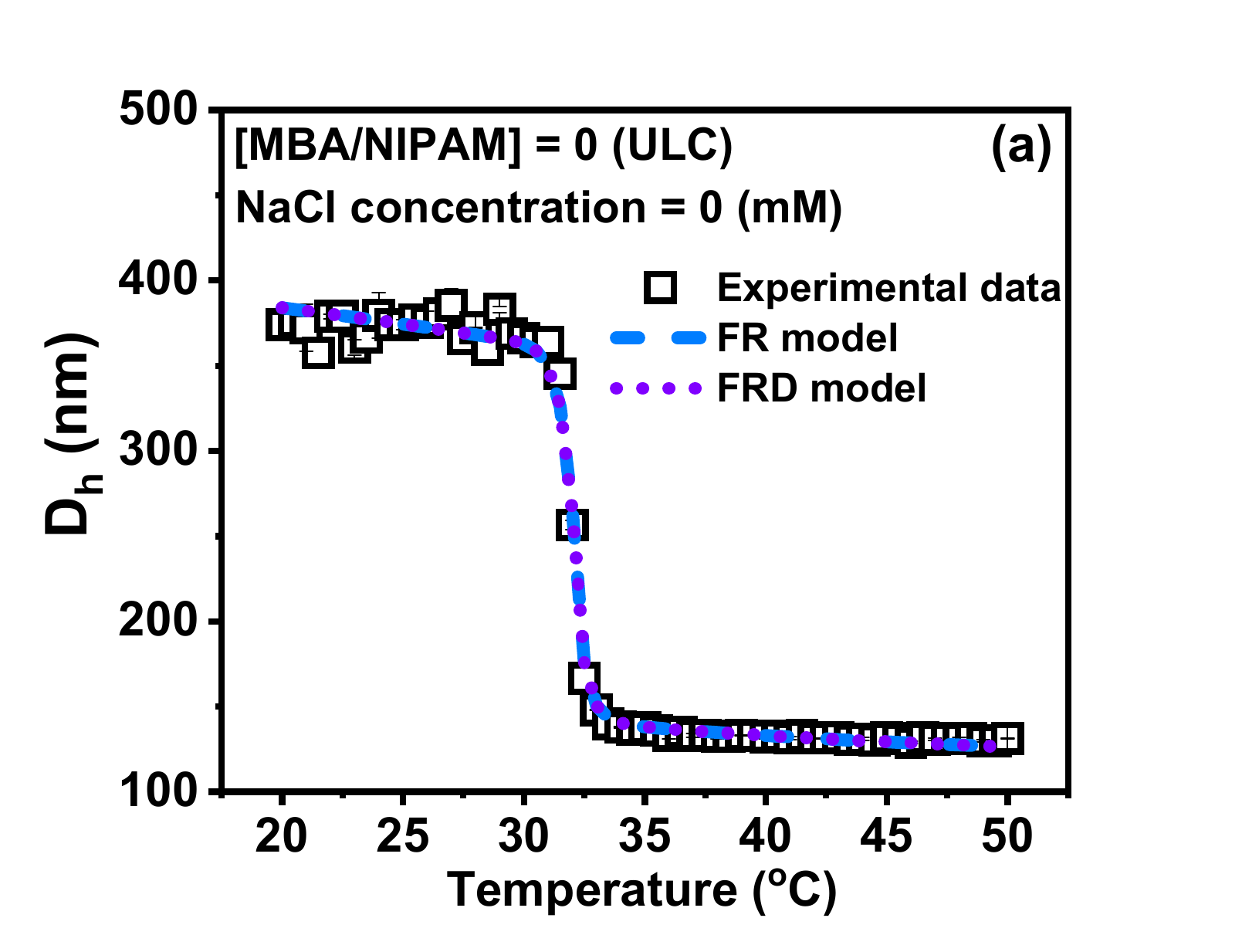}
  \end{minipage}
  \hfill
  \begin{minipage}{0.32\textwidth}
    \includegraphics[width=\textwidth]{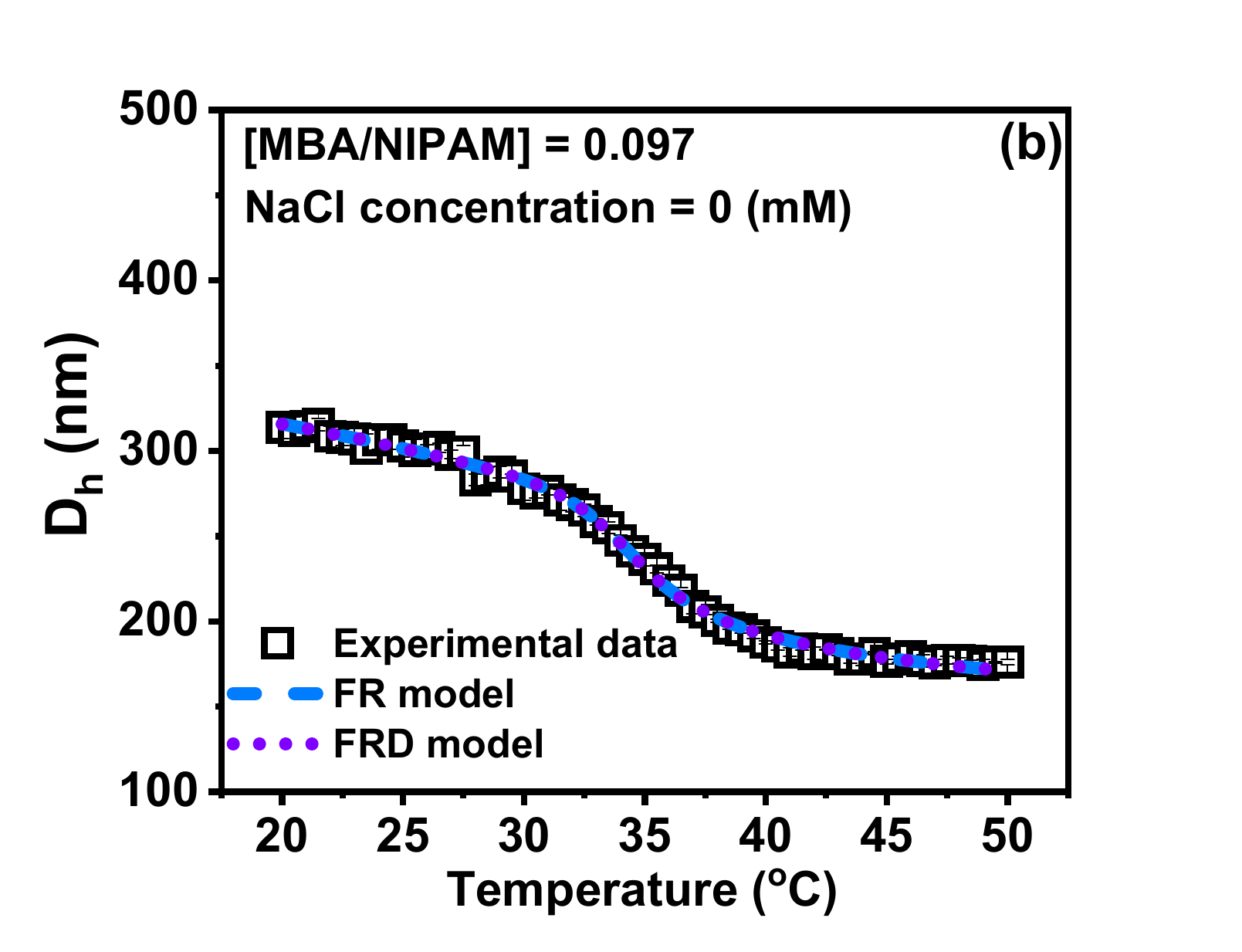}
  \end{minipage}
  \hfill
  \begin{minipage}{0.32\textwidth}
    \includegraphics[width=\textwidth]{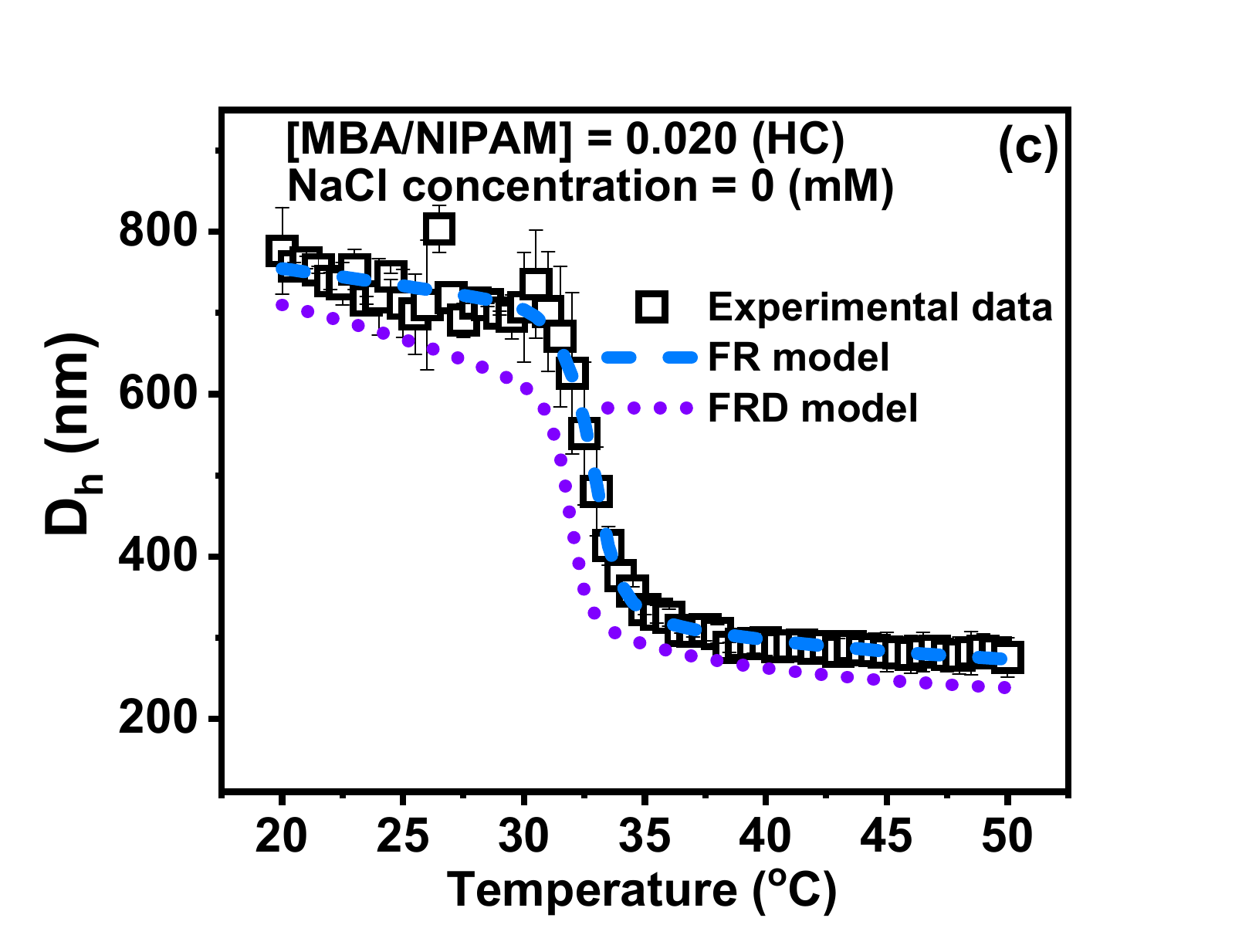}
  \end{minipage}

  \vspace{0.2cm} 

  \begin{minipage}{0.32\textwidth}
    \includegraphics[width=\textwidth]{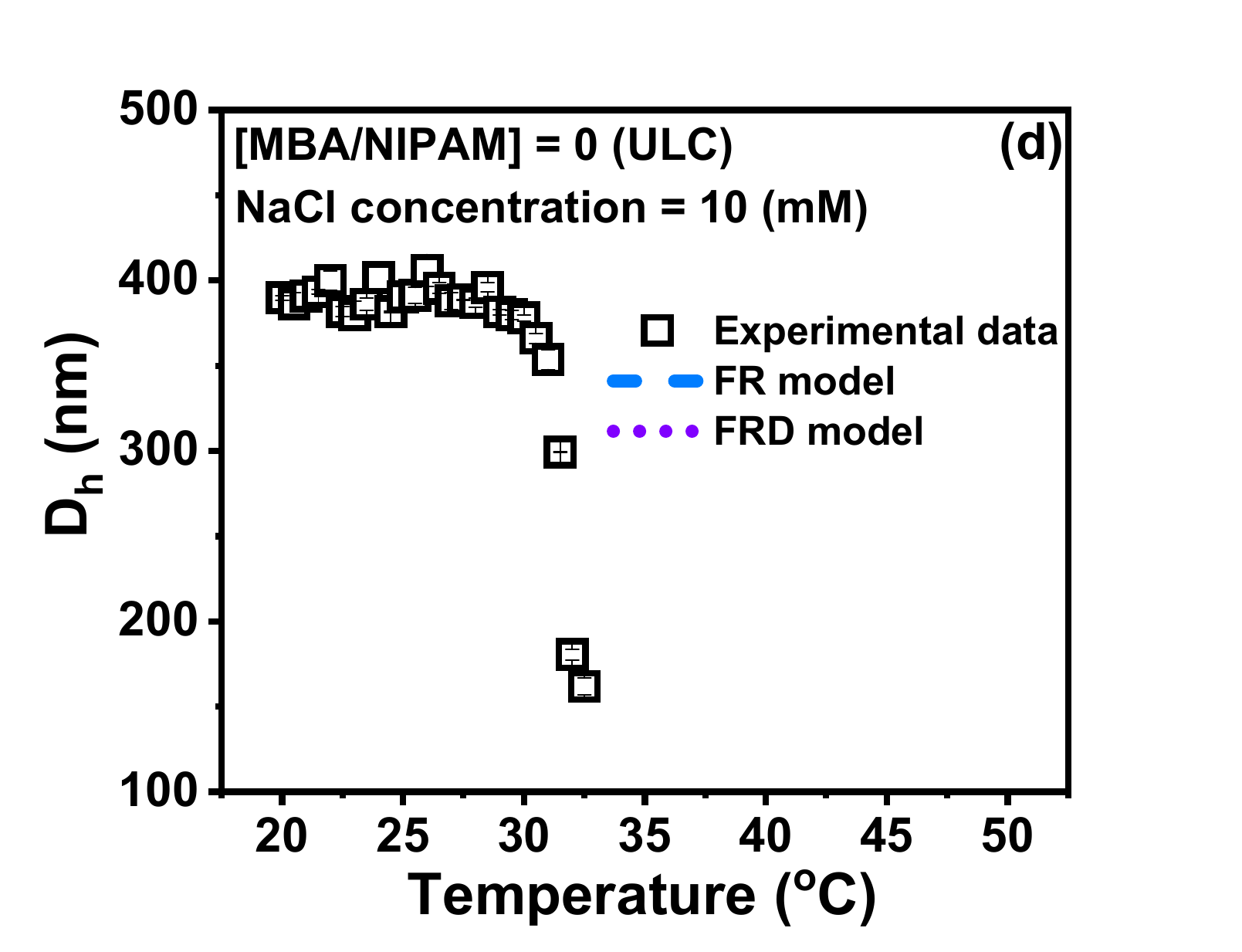}
  \end{minipage}
  \hfill
  \begin{minipage}{0.32\textwidth}
    \includegraphics[width=\textwidth]{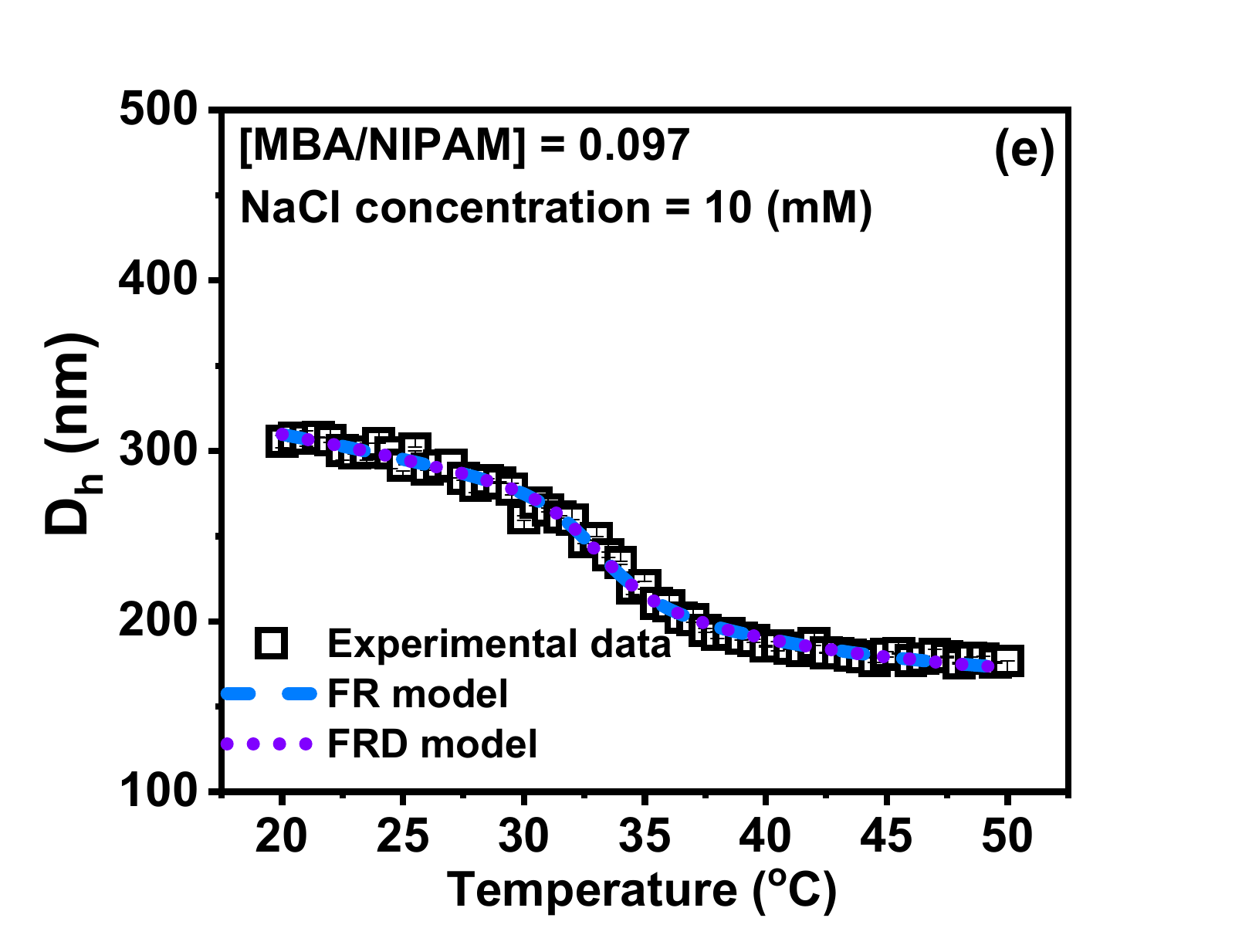}
  \end{minipage}
  \hfill
  \begin{minipage}{0.32\textwidth}
    \includegraphics[width=\textwidth]{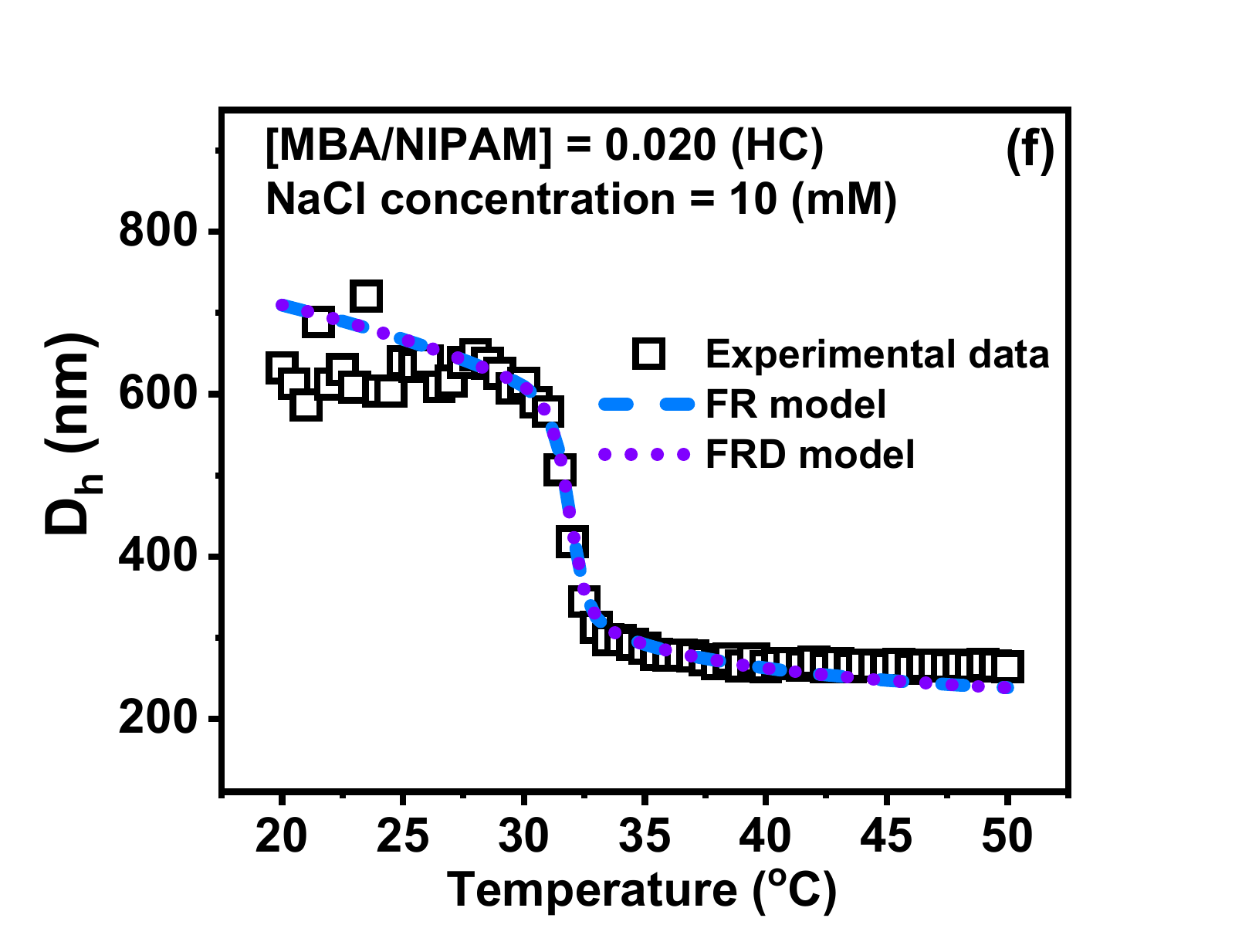}
  \end{minipage}

  \caption{The Flory-Rehner (FR) and Flory-Rehner-Donnan (FRD) models fit the experimental data: Temperature-dependent hydrodynamic diameter of pNIPAM microgels at different NaCl concentrations: (a, d) ULC microgel at 0 $mM$ and 10 $mM$ $NaCl$ respectively; (b, e) highly crosslinked microgel at 0 $mM$ and 10 $mM$ $NaCl$ respectively; (c, f) HC microgel at 0 $mM$ and 10 $mM$ $NaCl$ respectively. Symbols: open black square (experimental data), blue dashed line (Flory-Rehner model fit), and violet dotted line (Flory-Rehner-Donnan model fit).}
  \label{fig7}
\end{figure*}

\begin{table*}[t]
\centering
\begin{tabular}{|l|c|c|c|c|c|}
\hline
\textbf{Microgel} & \textbf{Theory} & \textbf{Nseg} & \textbf{$\phi_0$} & \textbf{$\gamma$ } & \textbf{$ (\chi^2)$} \\ \hline
\multirow{2}{*}{[MBA/NIPAM] = 0 (ULC)} & \textbf{FR} & 904.11 & 0.5583 & 25.00 & 9.39 \\ \cline{2-6}  
 & \textbf{FRD} & 911.89 & 0.5583 & 25.00 & 9.38 \\ \hline
\multirow{2}{*}{[MBA/NIPAM] = 0.007} & \textbf{FR} & 412.83 & 0.6795 & 14.03 & 10.36 \\ \cline{2-6}  
 & \textbf{FRD} & 412.62 & 0.6799 & 14.04 & 10.36 \\ \hline
\multirow{2}{*}{[MBA/NIPAM] = 0.020 (HC)} & \textbf{FR} & 621.19 & 0.5167 & 15.28 & 22.52 \\ \cline{2-6}  
 & \textbf{FRD} & 622.19 & 0.5166 & 15.24 & 22.52 \\ \hline
\multirow{2}{*}{[MBA/NIPAM] = 0.097} & \textbf{FR} & 36.69 & 0.6526 & 10.00 & 3.02 \\ \cline{2-6}  
 & \textbf{FRD} & 36.67 & 0.6528 & 10.00 & 3.02 \\ \hline
\end{tabular}
\caption{Critical parameters were obtained from Flory-Rehner (FR) and Flory-Rehner-Donnan (FRD) models fits to temperature-dependent hydrodynamic diameter data for distinct pNIPAM microgel formulations at 0 $mM$ $NaCl$}
\label{tab:microgel_data}
\end{table*}

We examined the applicability of the theoretical framework, specifically the Flory-Rehner and Flory-Rehner-Donnan models, to our experimental data sets. As shown in the Fig. \ref{fig7}, these models provide an excellent fit for the highly crosslinked (Fig. \ref{fig6} b,d) and homogeneous architectures (Fig. \ref{fig6} c,e) in both pure water and 10 $mM$ $NaCl$. A key observation is that, for our experimental data, the Flory-Rehner and Flory-Rehner-Donnan models yield identical results. This is attributed to the low intrinsic charge density that arises from the initiator residues; the Donnan term, though theoretically present, is not a dominant driver of the volume phase transition. The thermodynamics are primarily governed by the balance of mixing and elastic terms (Eq. \eqref{eq:flory_rehner}). This effectively justifies the use of the simpler Flory-Rehner for nonfunctionalized pNIPAM systems. 

ULC microgels presented a unique challenge because of the  the onset of flocculation at 10 $mM$ $NaCl$, which made the model fitting impossible (Fig. \ref{fig6}(d)). This is attributed to the fact that the relationship between polymer volume fraction and microgel size in the collapsed state could not be established (Eq. \eqref{eq:phi_relation}), as flocculation occurred before reaching the collapse state, making it impossible to determine the indivdual microgel particle size in the collapsed state at 10 $mM$ $NaCl$ concentration and, subsequently, the volume fraction. For ULC microgels, potentially there is a lack of a robust elastic or affine network, which allows the particles to deform and flocculate before reaching an equilibrium collapsed state; this violates the mean-field assumption of the Flory-Rehner model. This is a fundamental limitation of the theory when applied to non-affine structures. This finding supports the  argument of  Lopez and Richtering \cite{lopez2017does}, that the classical assumptions of the affine network break down when the crosslinking density is sufficiently low. To conclude our theoretical discussion, we examine the physical consistency of the parameters extracted from our fits at 0 $mM$ $NaCl$(Table. \ref{tab:microgel_data}). The most critical parameter is the $N_{\text{Seg}}$, that indicates the chain length between crosslinks. As expected $N_{\text{Seg}}$ decreases sharply as the crosslinking density increased, dropping from 904 to 36 (Table. \ref{tab:microgel_data}). This direct correlation with our synthesis recipes confirms that the model is capturing the actual network topology as reported by Friesen et al\cite{friesen2022modified}. In addition to that,  the $\phi_0$ values for the microgel range between 0.51 and 0.67, which is between the  debated values of $\phi_0$ existing in the literature\cite{lopez2017does,leite2018smart}. It can be inferred that $\phi_0$ is not a universal constant but a system-specific parameter that depends on the network topology of the microgel. Finally, regarding the quality of the fit, it is found that the highly crosslinked microgels yielded the lowest $(chi)^{2}$ values. This suggests that the classical Flory-Rehner model assumptions of a stable affine network are most accurate when the microgel has a robust elastic backbone. The elevated $chi^2$ values for the ULC and HC architectures reflect the non-affine nature and enhanced thermal fluctuations inherent to low crosslinking density networks, which depart from the idealized elastic framework of the Flory-Rehner model. These findings reinforce that the model is mathematically a powerful tool, but it requires physical grounding in the microgel architecture to be truly predictive.

\section{Conclusion}

This study systematically elucidates the interplay of network crosslinking distribution, density, the presence or absence of crosslinkers, thermodynamic states, and the external ionic environment in defining the fundamental thermoresponsive properties and colloidal stability of non-functionalized pNIPAM microgels. Our findings indicate that the volume phase transition temperature (VPTT) is primarily governed by crosslinker concentration and ionic strength. Interestingly, while the VPTT increases with crosslinker density, the magnitude of VPTT reduction upon salt addition is most pronounced in highly crosslinked systems, where elastic contributions become dominant. Notably, the spatial distribution of crosslinks and even the total absence of a crosslinker (ULC) were found to have a negligible impact on the sensitivity of the VPTT to ionic strength, with ULC and homogeneously crosslinked (HC) microgels exhibiting shifts comparable to those with low MBA concentrations.

Regarding the salt tolerance, the network architecture plays a substantial role. Core-corona microgels maintain superior structural integrity in saline media because of their dense elastic network provides the necessary restorative forces to counteract osmotic pressure drops. Conversely, the absence of MBA crosslinks in ULC microgels results in a negligible elastic network, leading to a unique regime of anomalous swelling at low temperatures followed by complete collapse during and beyond VPT. HC microgels, potentially due to their homogeneous distribution of crosslinking in the network and hence the absence of highly crosslinked core, make it susceptible to osmotic deswelling and significant salting-out effects during the entire temperature range. 

These architectural differences extend to thermoreversibility and colloidal stability. Under high ionic stress ($100$ $mM$ $NaCl$), the lack of an elastic framework leads to irreversible flocculation; soft microgels at these conditions lose reversibility. The presence of MBA crosslinks is critical for maintaining this reversibility, as ULC microgels becomes sticky even at lower salt concentrations compared to MBA-containing variants. Beyond equilibrium swelling, our analysis of flocculation kinetics at extreme salinity ($1000$ $mM$ $NaCl$) reveals that network topology dictates the pathway to flocculation. Interestingly, ULC and core-corona microgels with low core crosslinking exhibit rapid, irreversible flocculation, whereas homogeneously crosslinked and core-corona microgels with highly crosslinked microgels demonstrate significantly slower flocculation rates. 

Finally, our evaluation of the Flory-Rehner and Flory-Rehner-Donnan models against this extensive 22-batch library yields a critical insight: the identical performance of both models underscores that for non-functionalized pNIPAM systems, the chemical salting-out effect on the polymer backbone is the dominant thermodynamic driver. In the absence of ionic comonomers, the Donnan term contribution remains negligible. These findings reinforce that while these models are mathematically powerful tools, their predictive accuracy depends upon physical grounding in the specific microgel architecture. This work provides a robust predictive framework for designing salt-tolerant thermoresponsive colloids. Specifically, the core-corona architecture is identified as a superior template for applications demanding structural robustness in complex ionic environments, such as wastewater treatment, controlled drug release, or smart actuation.

\begin{acknowledgments}
The authors appreciate the invaluable support of the Soft Matter Group (SMG) lab members throughout this project. Their assistance and collaborative spirit contributed significantly to the research. The authors acknowledge the financial support provided by the Science and Engineering Research Board (SERB) for the project (No. SRG/2021/000779), and ARJ specifically acknowledges the seed grant (No. SG99) provided by IIT Hyderabad. The funding enabled us to acquire the necessary materials and equipment, allowing us to conduct this research.  The authors thank Dr. Mahesh Ganesan for suggestions on DLS flocculation kinetics. ARJ dedicates this manuscript to Prof. Jan Mewis.

\end{acknowledgments}

\section{DATA AVAILABILITY STATEMENT}

All data used in the figures of the main text and the supplemental material are available on request.

\clearpage
\section{SUPPLEMENTARY MATERIAL}

\subsection{\label{app:subsec}Synthesis protocol}
\subsubsection{One-pot synthesis}
pNIPAM microgels with core–corona structure and varying crosslinker concentrations were synthesized via conventional one-pot synthesis. For a representative synthesis with an [MBA/NIPAM] mole ratio of 0.007, a monomer solution was prepared by dissolving \SI{2310}{\milli\gram} of NIPAM, \SI{24.4}{\milli\gram} of MBA, and \SI{24.2}{\milli\gram} of SDS in \SI{225}{\milli\liter} of  water. The water used (type 1 quality) for the entire synthesis and sample preparation was filtered through \SI{0.4}{\micro\metre} syringe filters to minimize contaminants. The monomer solution was stirred at \SI{1000}{rpm} for \SI{15}{\minute} to ensure uniform mixing. A three-neck flask equipped with a reflux condenser (connected to a chiller set at \SI{10}{\degreeCelsius}) and a nitrogen inlet was submerged in a silicon oil bath (10 cSt) on a hot plate stirrer heated to \SI{85}{\degreeCelsius} and set at \SI{1000}{rpm} stirring. Once the oil bath was reached \SI{75}{\degreeCelsius}, the monomer solution was transferred to the three-neck flask. Following transfer, the reactor was purged with nitrogen for \SI{30}{\minute} to make an oxygen-free atmosphere. An initiator solution of \SI{90}{\milli\gram} of KPS dissolved in \SI{25}{\milli\liter} of water was added once the oil bath reached \SI{85}{\degreeCelsius} to initiate polymerization and prepare anionic (negatively charged) pNIPAM microgels. The reaction mixture transformed from translucent to turbid white within minutes, indicating active polymerization. The reaction was carried out for \SI{4}{\hour} at \SI{85}{\degreeCelsius} and \SI{1000}{rpm} stirring using the hot plate stirrer. The chiller was set at \SI{10}{\degreeCelsius} and nitrogen was purged into the three-neck flask throughout the reaction. After which the mixture was allowed to cool overnight. The product was purified through centrifugation (a minimum of four cycles at \SI{7400}{rpm} and \SI{37}{\degreeCelsius} for \SI{90}{\minute} per cycle). Between cycles, the supernatant was discarded, and the sedimented microgel was resuspended in water. Following purification, \SI{2}{\milli\mole} of sodium azide (NaN$_3$) was added as a preservative to avoid bacterial growth.

Additional batches with varying crosslinker concentrations were synthesized using the identical methodology while maintaining constant quantities of all other reagents. The [MBA/NIPAM] mole ratios employed were 0.031, 0.038, 0.058, 0.077, and 0.097, respectively (with the [KPS/NIPAM] mole ratio = 0.016 and the [SDS/NIPAM] mole ratio = 0.004 held constant). Replicates for each crosslinker concentration were synthesized to ensure consistency in microgel characteristics.

\textbf{Ultra-low crosslinked (ULC) microgels} were synthesized following the same procedure but without the addition of the MBA crosslinker, maintaining a constant [KPS/NIPAM] mole ratio of 0.016 and [SDS/NIPAM] mole ratio of 0.004. Two replicates for the ULC microgels were synthesized to ensure consistency in microgel characteristics.

\subsubsection{Semi-batch synthesis}
The synthesis of pNIPAM microgels with homogeneous crosslinking density ([MBA/NIPAM] mole ratio = 0.020) was done via semi-batch synthesis. The reaction setup consisted of a three-neck flask equipped with a reflux condenser (connected to a chiller set at \SI{10}{\degreeCelsius}), a nitrogen inlet, and a syringe pump connected to the third neck for controlled monomer-crosslinker feed. The flask was submerged in a silicon oil bath (10 cSt) on a hot plate stirrer heated to \SI{85}{\degreeCelsius} and set at \SI{1000}{rpm}.

The monomer solution was prepared by dissolving \SI{0.9}{\gram} of NIPAM in \SI{55}{\milli\liter} of water. A feed solution was prepared by dissolving \SI{0.3}{\gram} of NIPAM and \SI{32}{\milli\gram} of MBA in \SI{6}{\milli\liter} of water, followed by degassing for \SI{30}{\minute}. Polymerization was initiated by adding \SI{55}{\milli\gram} of KPS dissolved in \SI{1}{\milli\liter} of water to the initial monomer solution. Four minutes after KPS addition, the solution became milky, indicating the active polymerization. The second stage commenced with controlled feeding of the monomer-crosslinker solution at a rate of \SI{100}{\micro\liter\per\minute} using the syringe pump. Upon completion of the feed solution, the reaction was terminated by quickly cooling the flask in an ice bath, resulting in a turbid microgel suspension. The product was purified using the same centrifuge protocol as described for the one-pot synthesis. Replicates for this crosslinker concentration were synthesized to ensure consistency in microgel characteristics.

\newpage
\subsection{Chemical Structures and Reaction Mechanisms}

\begin{figure}[htbp]
\centering
\includegraphics[scale=0.4]{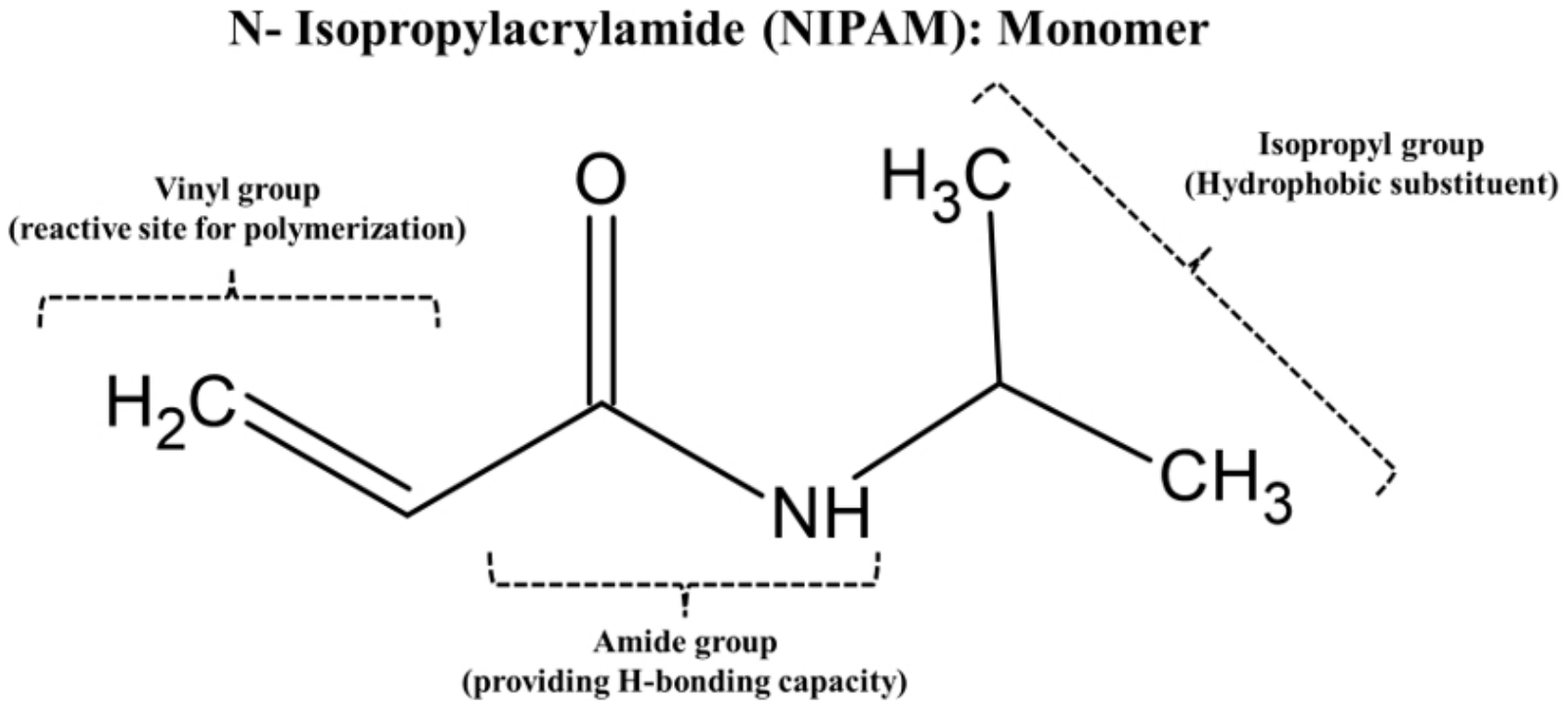}
\includegraphics[scale=0.7]{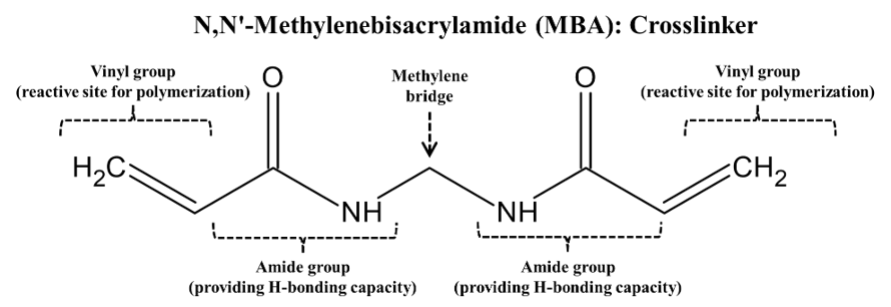}
\caption{Chemical structures of NIPAM and MBA (referenced in Materials and Methods)}
\label{SI1}
\end{figure}

\begin{figure*}[htbp]
\centering
\includegraphics[height=10cm]{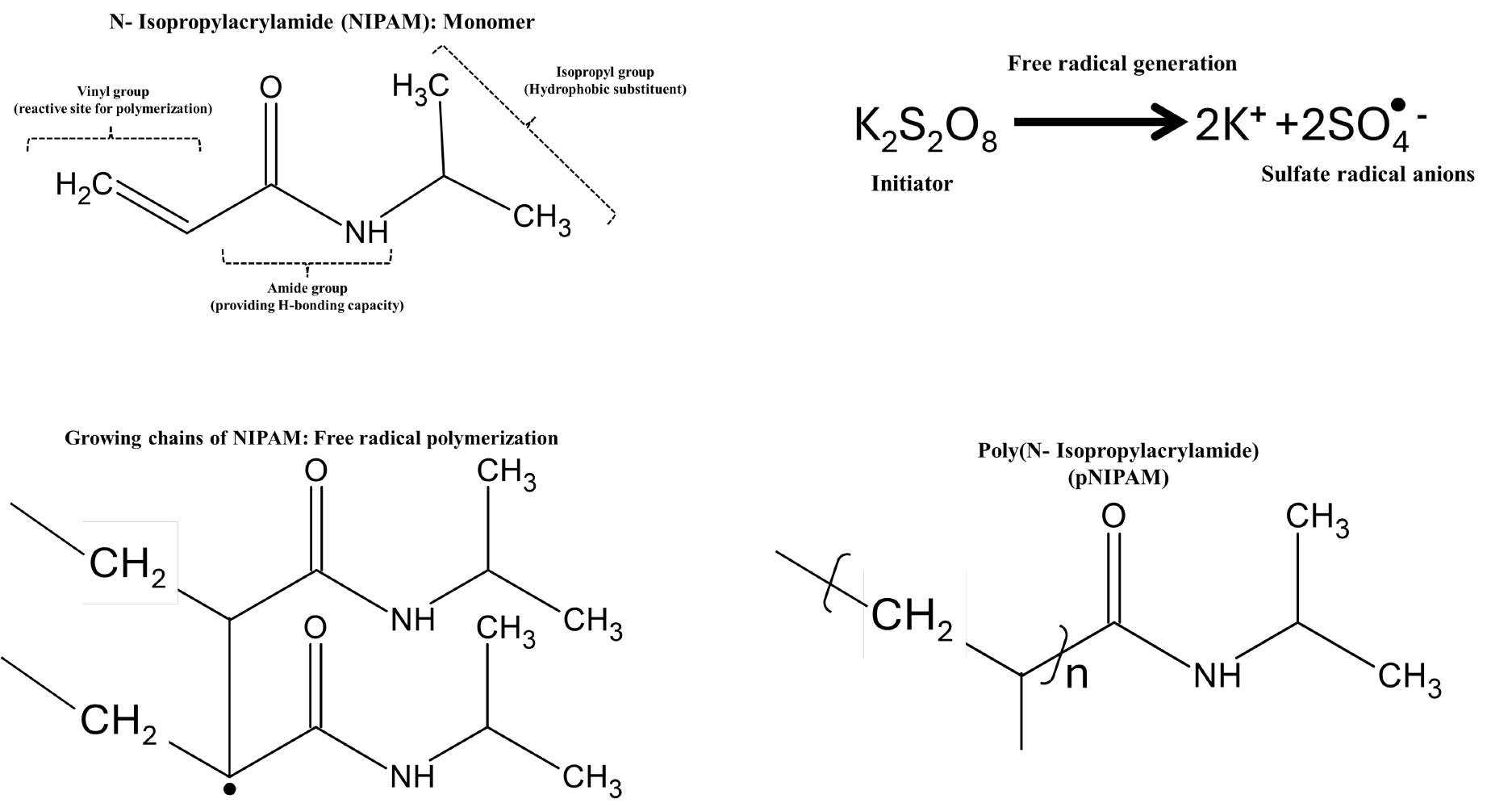}\hfill
\includegraphics[height=8cm]{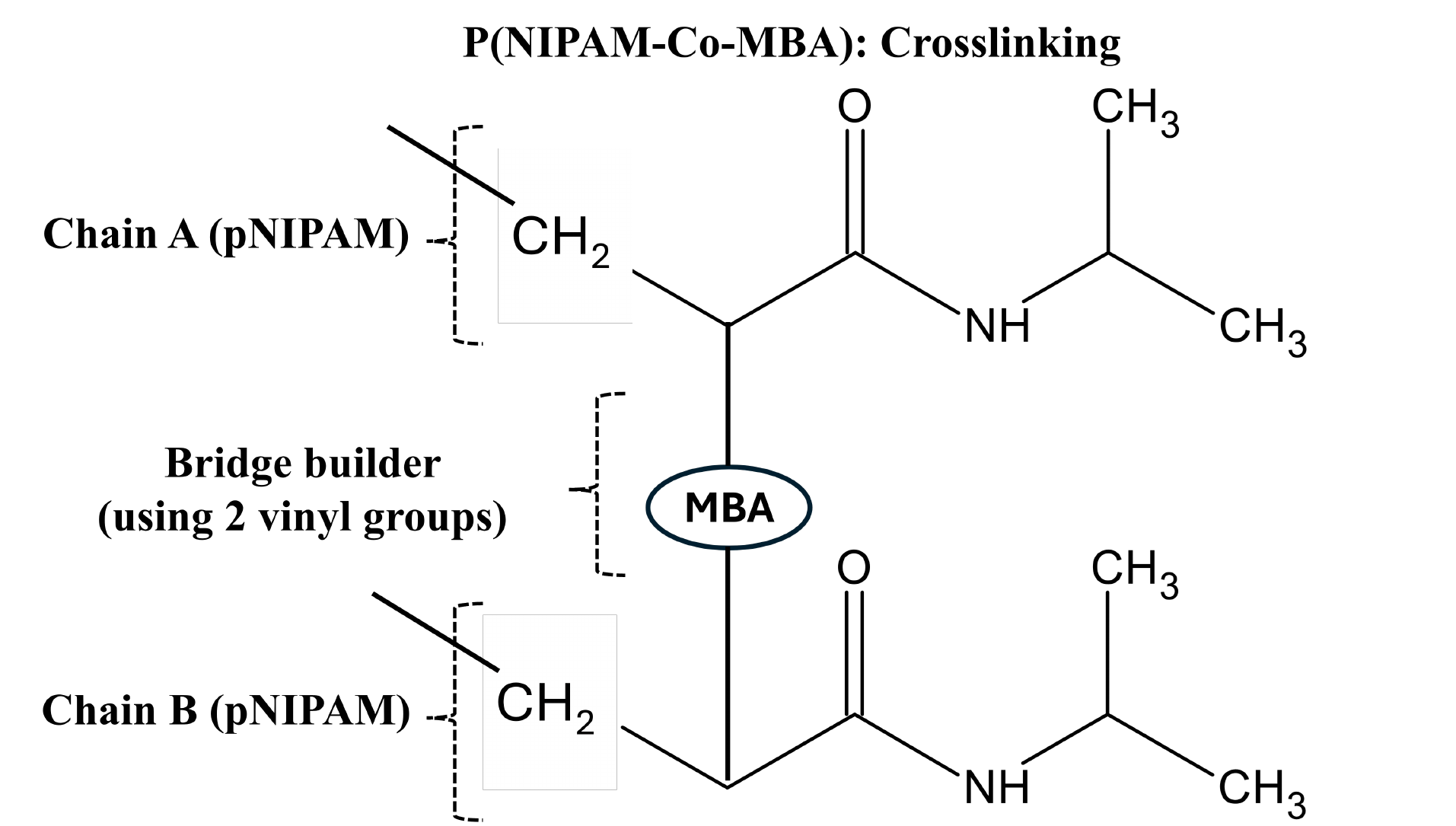}
\caption{Schematic representation of the free radical polymerization and crosslinking mechanism for core-corona microgels. (Referenced in Materials and Methods)}
\label{SI2}
\end{figure*}
\begin{figure*}[htbp]
\centering
\includegraphics[height=8cm]{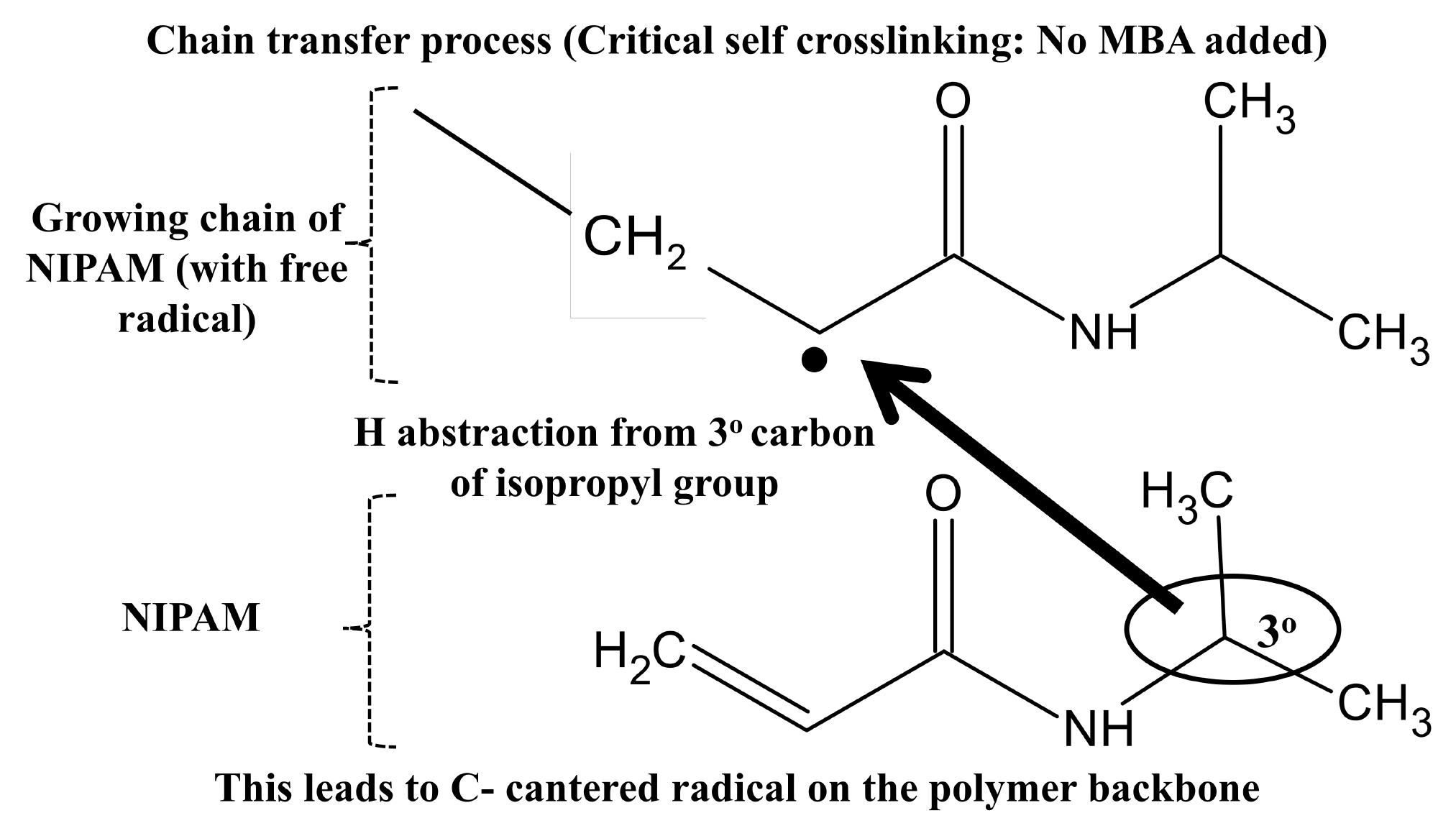}
\caption{Schematic representation of the self-crosslinking mechanism via chain transfer for ULC microgels. (Referenced in Materials and Methods)}
\label{SI10}
\end{figure*}

\FloatBarrier
\newpage

\subsection{Extended thermoresponsive data}  

\begin{figure}[H] 
\centering
\includegraphics[height=5.5cm]{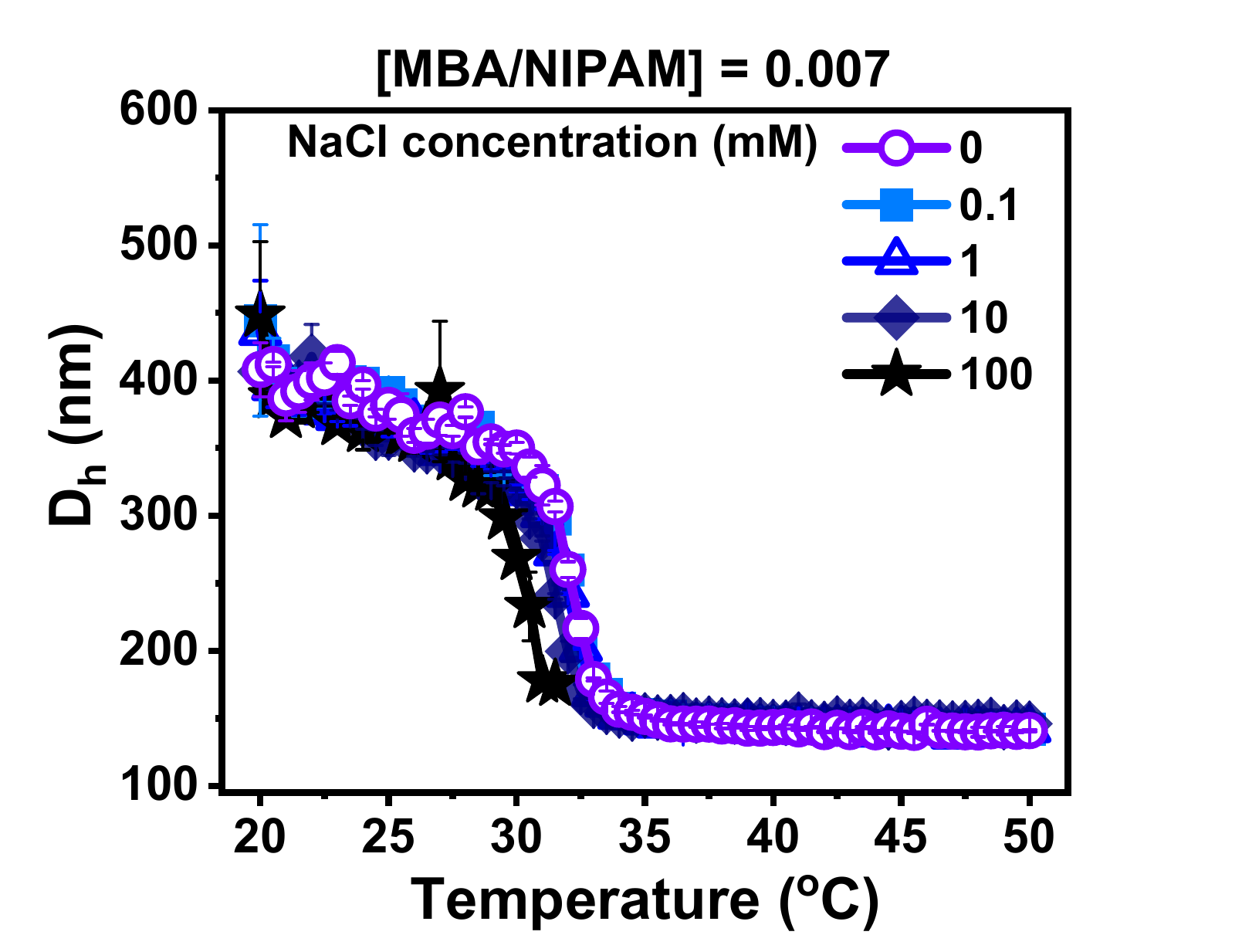}
\caption{Full temperature sweeps (\SI{20}{\celsius} to \SI{50}{\celsius}) of hydrodynamic diameter for microgel with [MBA/NIPAM] = 0.007 (Referenced in Materials and Methods)}
\label{SI:007}
\end{figure}

\begin{figure}[H] 
\centering
  \includegraphics[height=5.5cm]{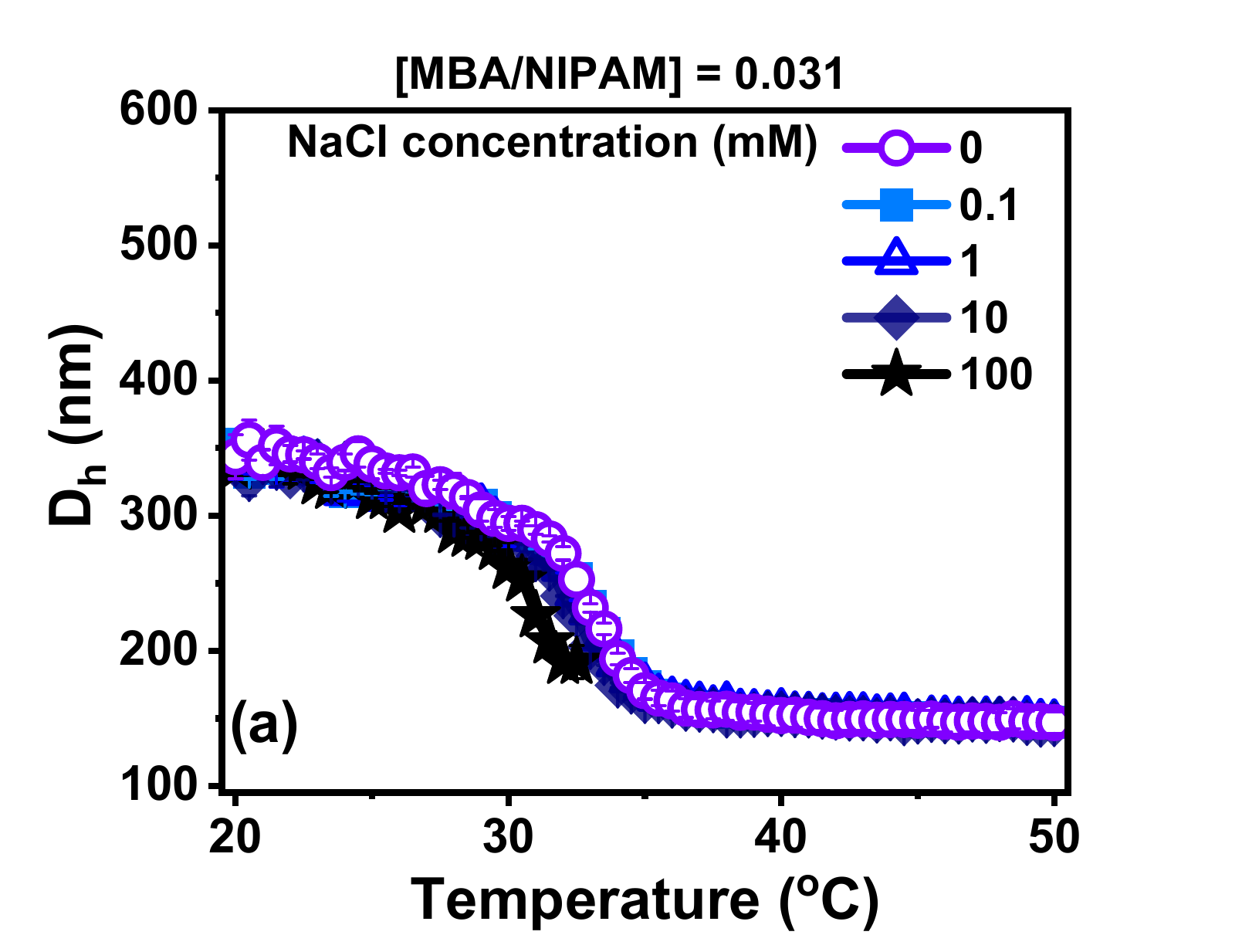}
  \caption{ Full temperature sweeps (\SI{20}{\celsius} to \SI{50}{\celsius}) of hydrodynamic diameter for microgel with [MBA/NIPAM] = 0.031] (Referenced in Materials and Methods)}
  \label{SI:0.031}
\end{figure}

\begin{figure}[H] 
\centering
  \includegraphics[height=5.5cm]{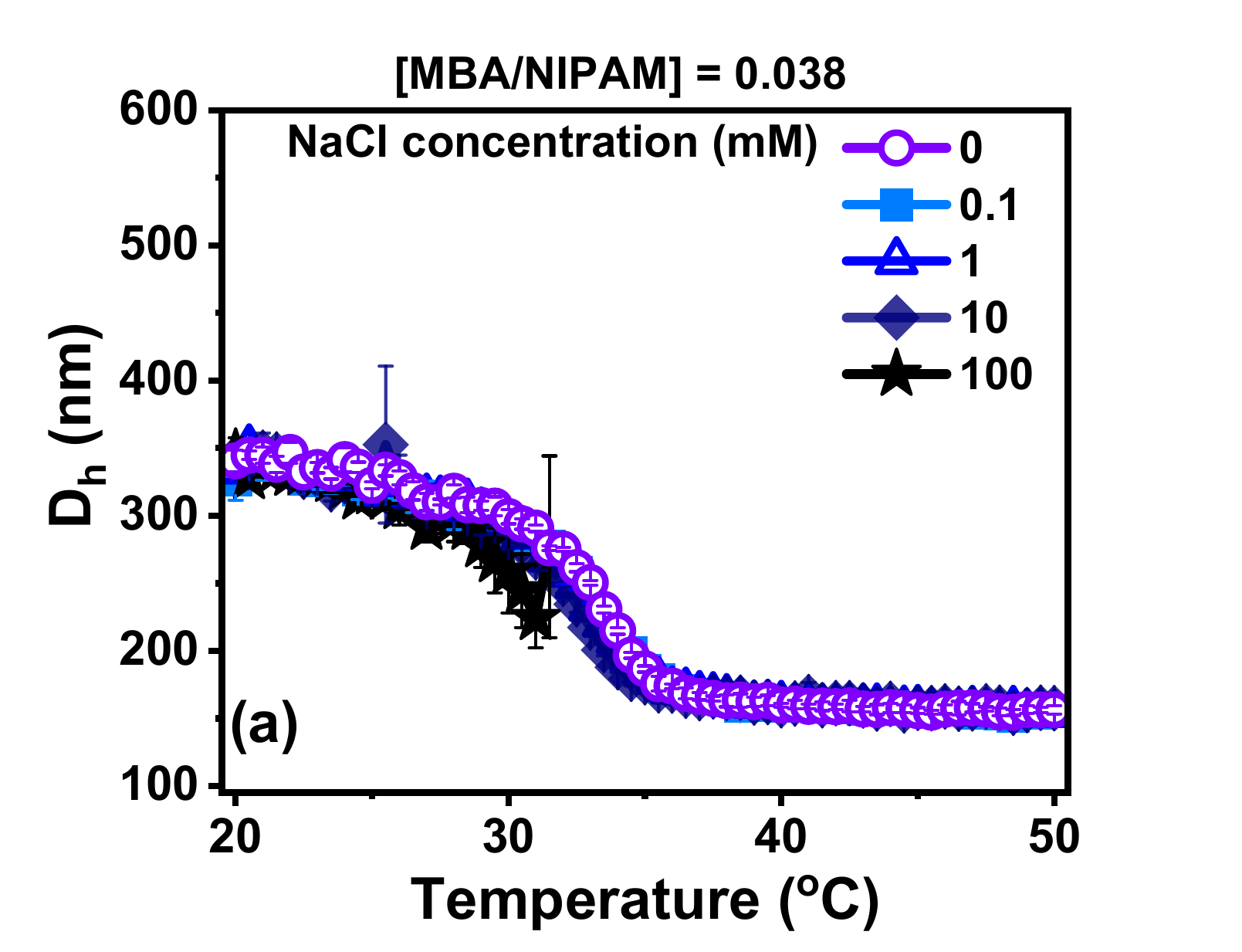}
  \caption{ Full temperature sweeps (\SI{20}{\celsius} to \SI{50}{\celsius}) of hydrodynamic diameter for microgel with [MBA/NIPAM] = 0.038] (Referenced in Materials and Methods)}
  \label{SI:0.038}
\end{figure}

\begin{figure}[H] 
\centering
  \includegraphics[height=5.5cm]{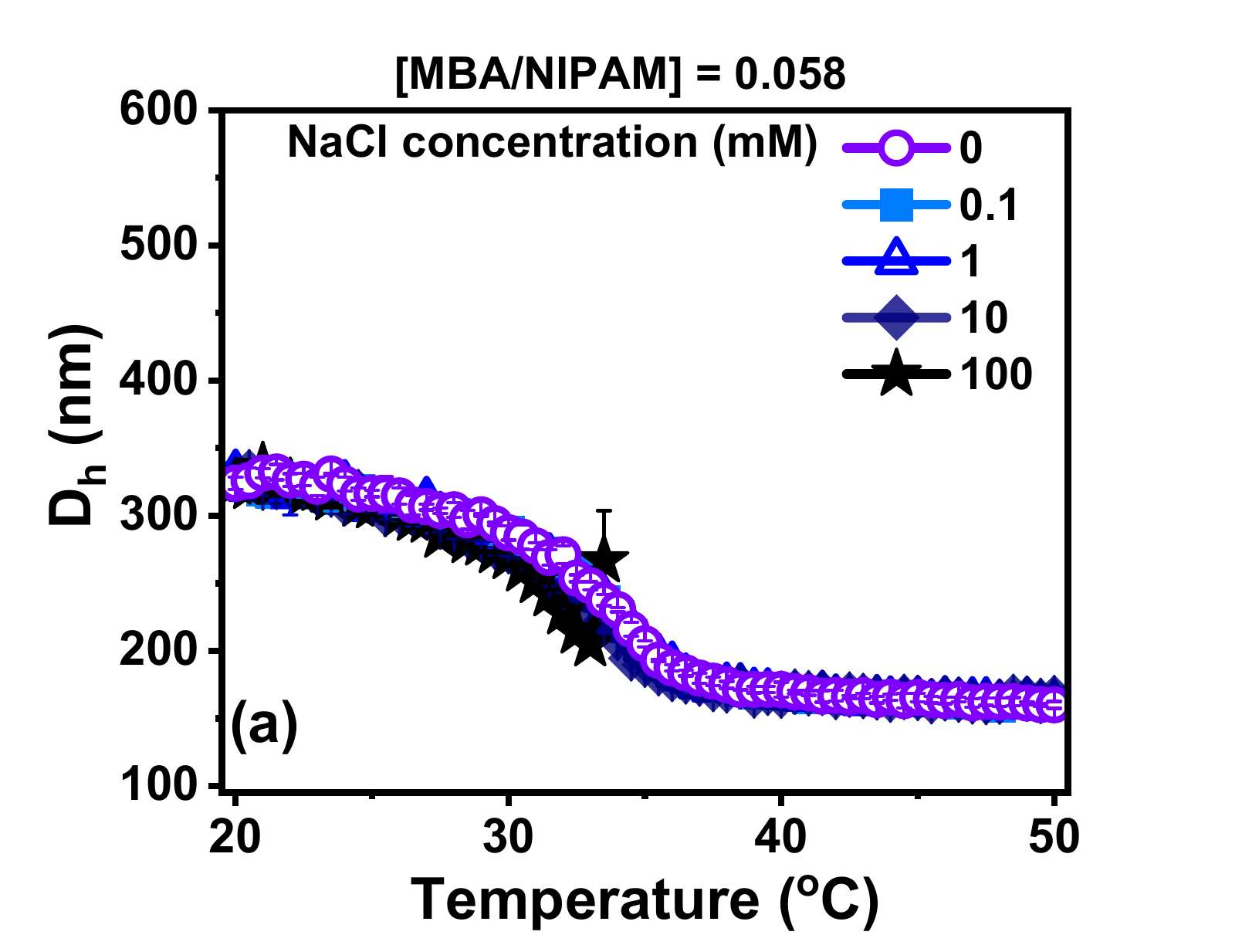}
  \caption{ Full temperature sweeps (\SI{20}{\celsius} to \SI{50}{\celsius}) of hydrodynamic diameter for microgel with [MBA/NIPAM] = 0.058] (Referenced in Materials and Methods)}
  \label{SI:0.058}
\end{figure}

\begin{figure}[H] 
\centering
  \includegraphics[height=5.5cm]{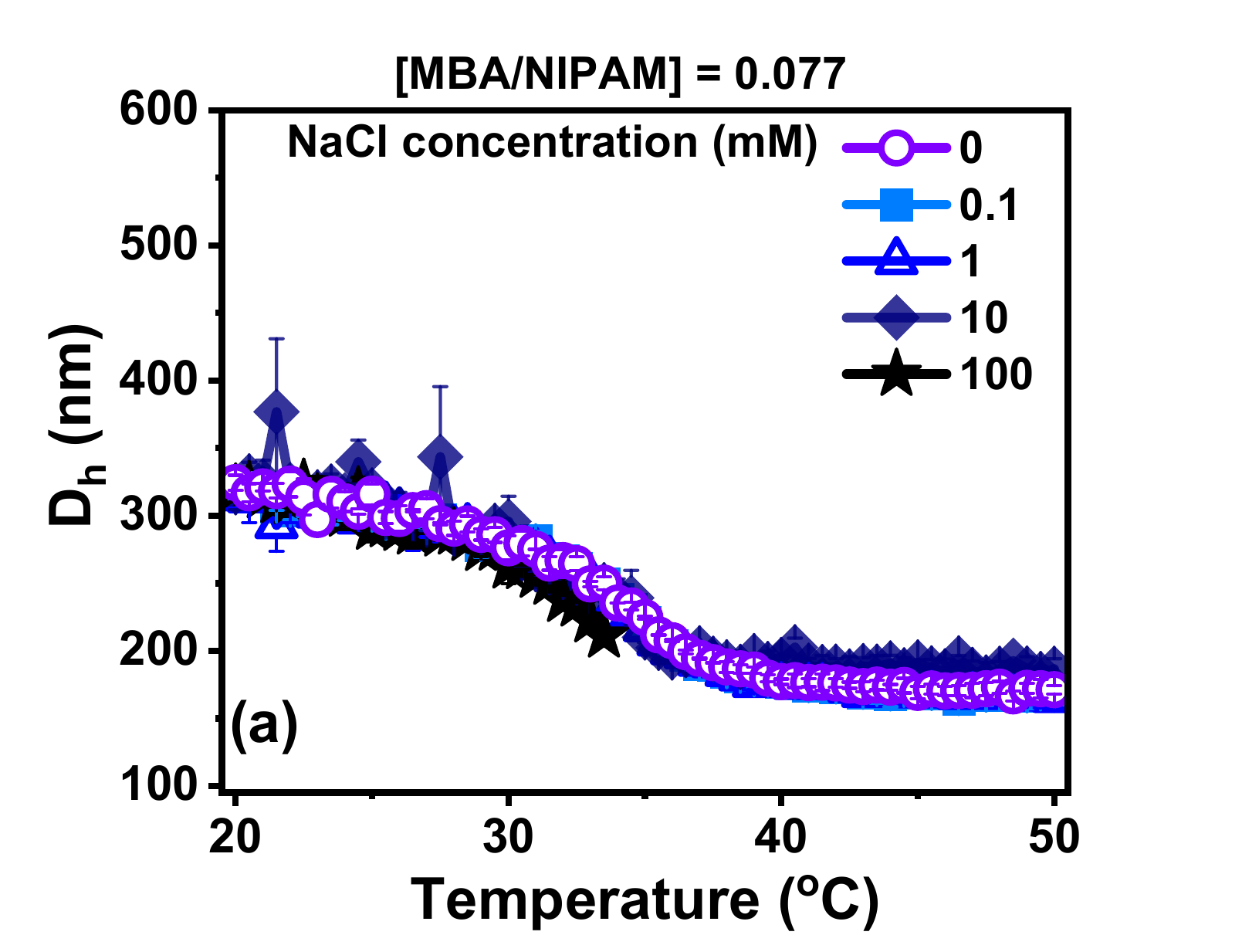}
  \caption{ Full temperature sweeps (\SI{20}{\celsius} to \SI{50}{\celsius}) of hydrodynamic diameter for microgel with [[MBA/NIPAM] = 0.077] (Referenced in Materials and Methods)}
  \label{SI:0.077}
\end{figure}

\FloatBarrier
\newpage
\subsection{Extended salt tolerance data}

\begin{figure}[H]
\centering
  \includegraphics[height=6cm]{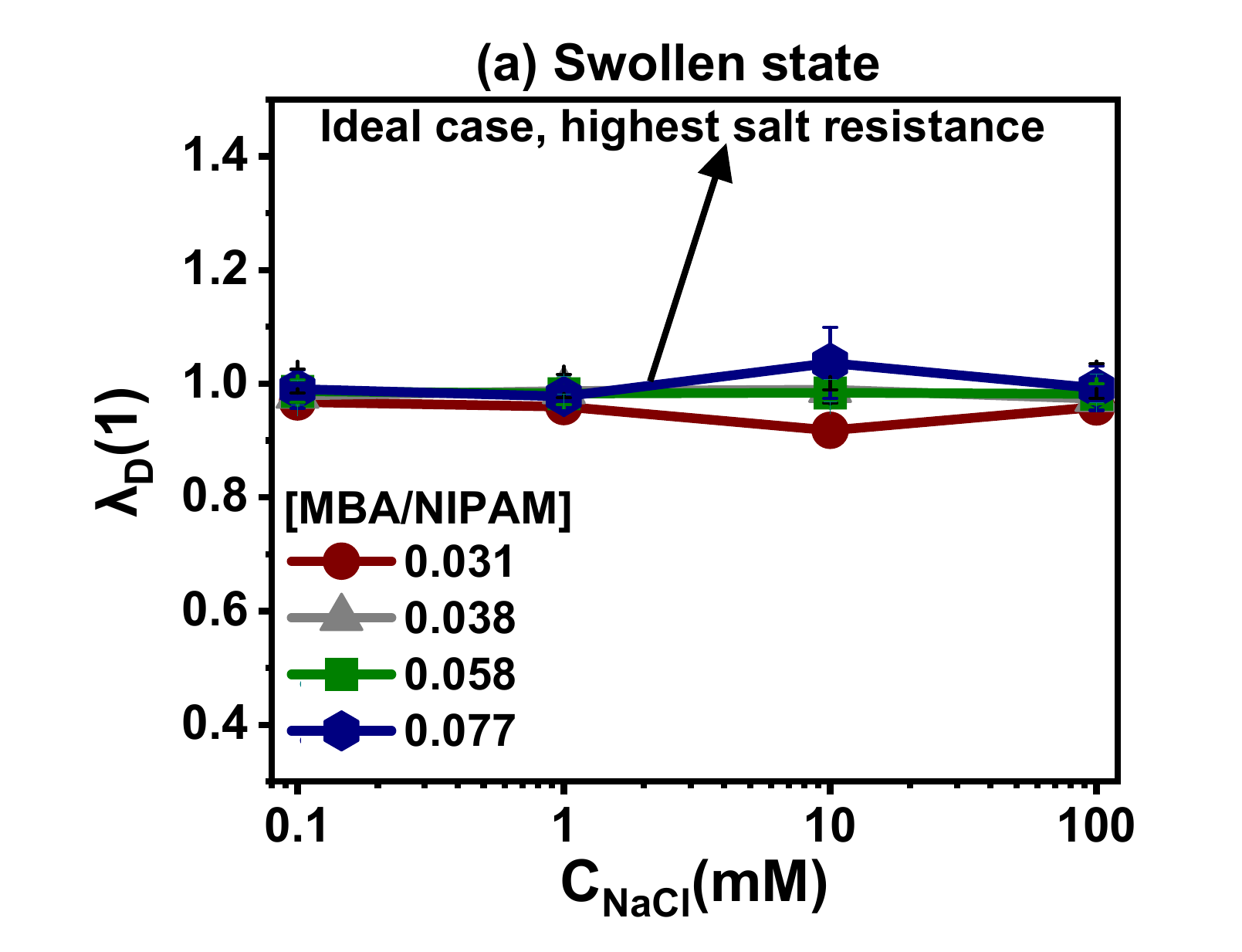}
  \includegraphics[height=6cm]{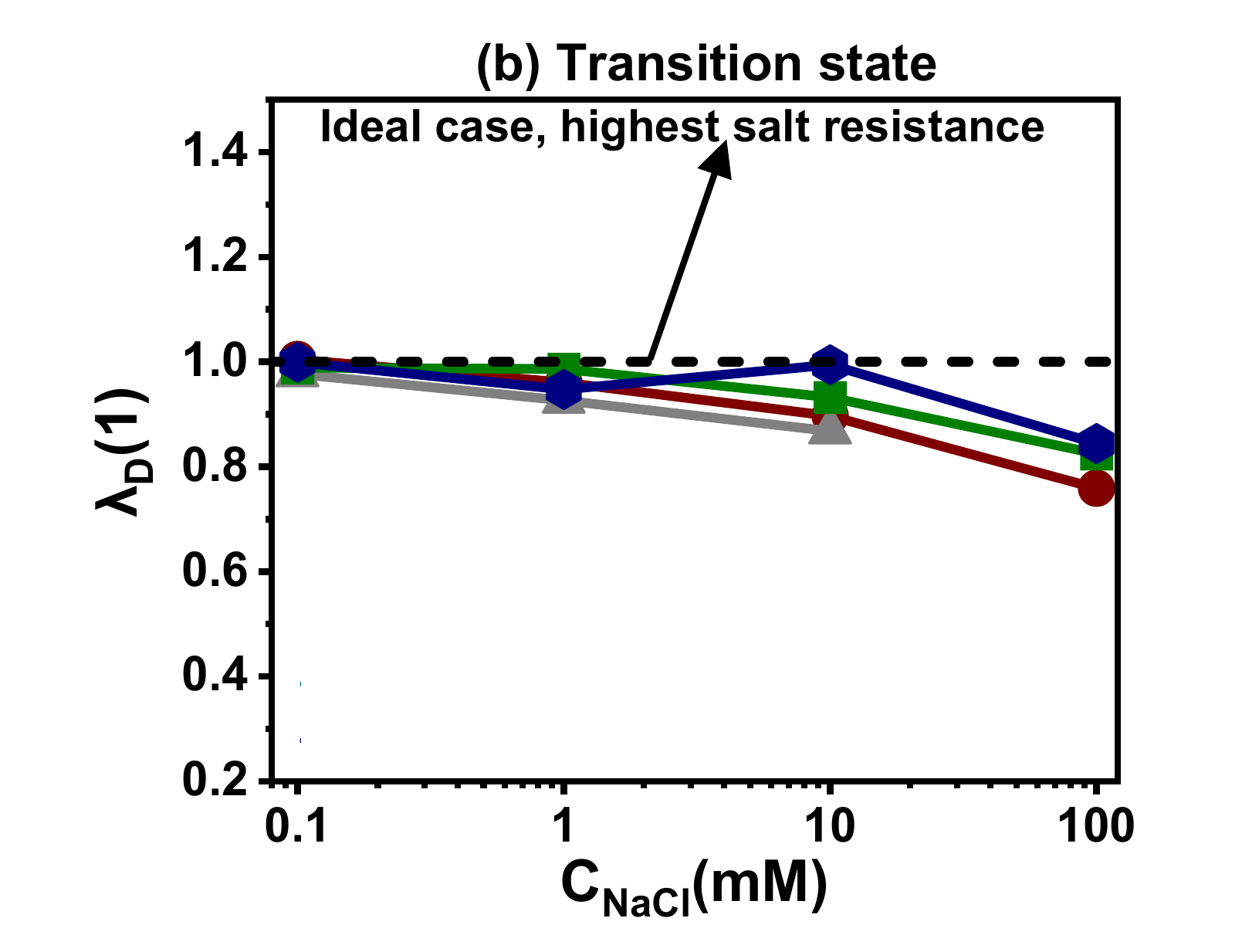}
  \includegraphics[height=6cm]{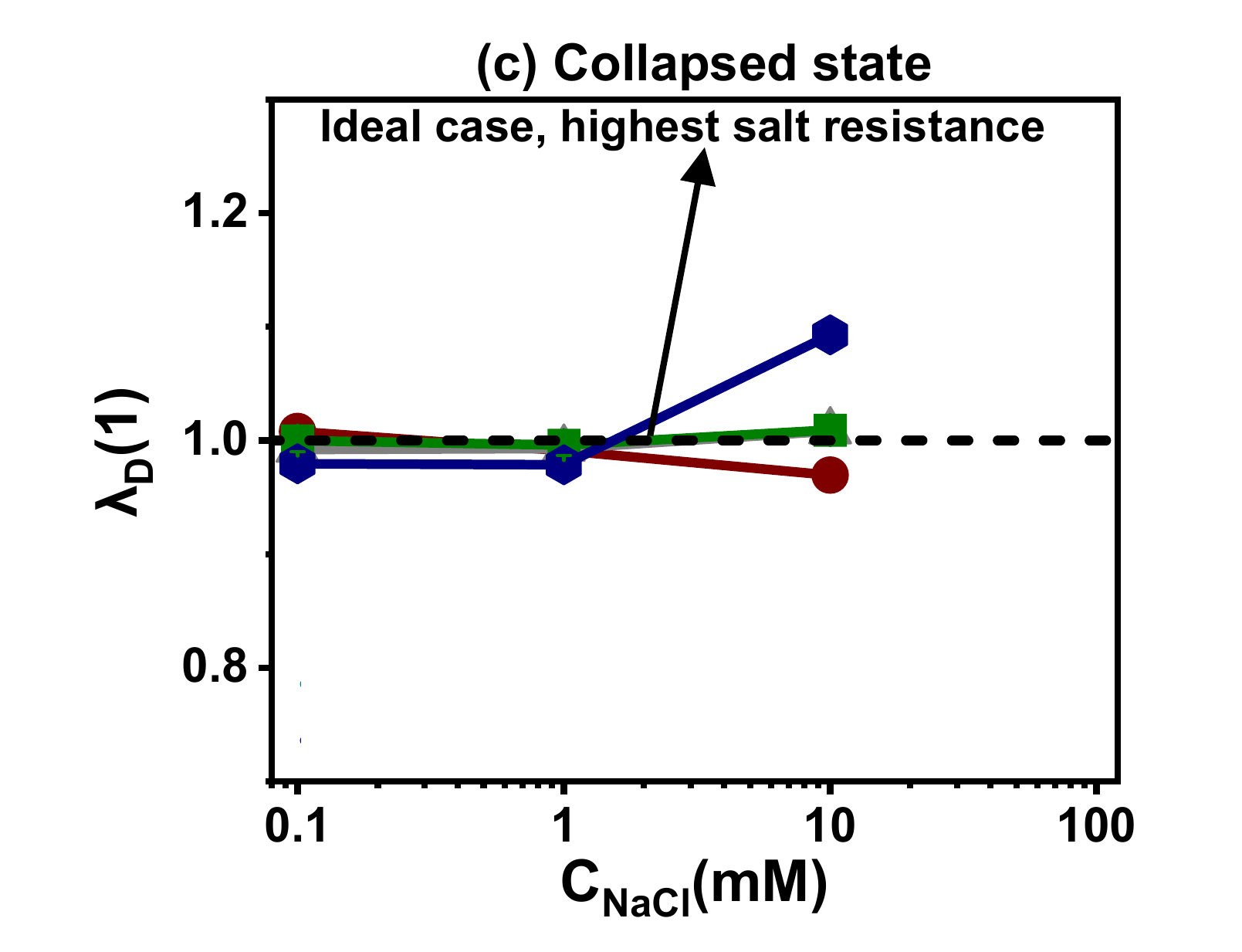}
  \caption{Normalized Size ($\lambda_D$) vs. NaCl concentration for the intermediate crosslinking density samples}
  \label{SI11}
\end{figure}

\FloatBarrier
\newpage

\subsection{Extended HI Data}
\begin{figure}[H]
\centering
  \includegraphics[height=6cm]{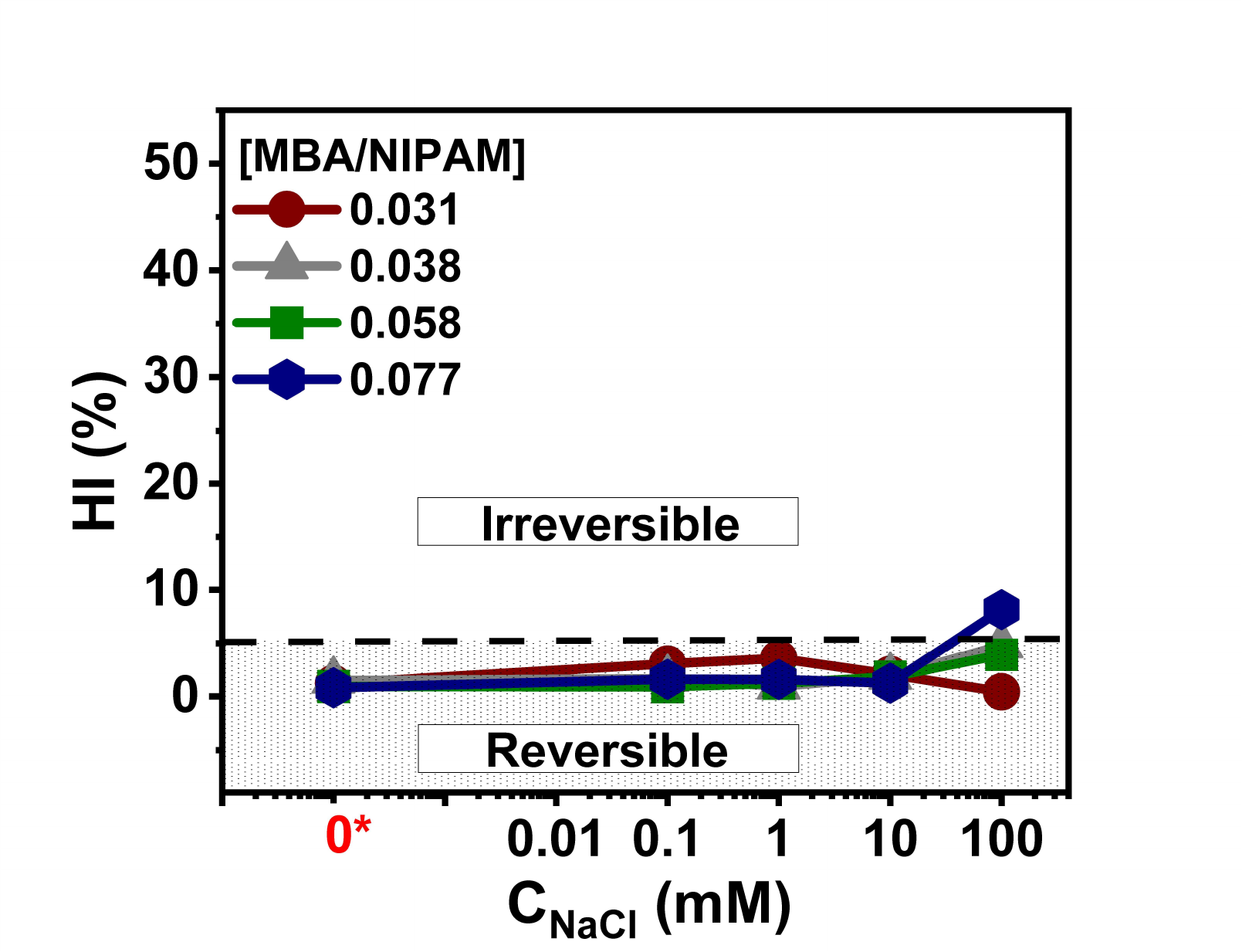}
  \caption{ Hysteresis index (HI) vs. NaCl concentration for the intermediate crosslinking density microgels}
  \label{SI12}
\end{figure}

\nocite{*}

\bibliography{apssamp}

\end{document}